\documentclass[rmp,aps,nofootinbib,twocolumn,letterpaper,floatfix]{revtex4}
\usepackage{graphics}
\usepackage{graphicx}
\usepackage{epsfig}
\usepackage{rotating}
\usepackage{lscape}
\usepackage{longtable}

\begin{document}
\title{Physical approaches to DNA sequencing and detection}

\author{Michael Zwolak}
\email{mpz@lanl.gov}
\affiliation{Physics Department, California Institute of Technology, Pasadena, California 91125}
\affiliation{Theoretical Division MS-B213, Los Alamos National Laboratory, Los Alamos, NM 87545}
\author{Massimiliano Di Ventra}
\email{diventra@physics.ucsd.edu}
\affiliation{Department of Physics, University of California, San Diego, La Jolla, CA 92093}

\begin{abstract}
With the continued improvement of sequencing technologies, the prospect of
genome-based medicine is now at the forefront of scientific research. To realize this potential, however,
we need a revolutionary sequencing method for the cost-effective and rapid interrogation
of individual genomes. This capability is likely to be provided by a {\it physical approach}
to probing DNA at the single nucleotide level. This is in sharp contrast to
current techniques and instruments which probe, through chemical elongation, electrophoresis,
and optical detection, length differences and terminating bases of strands of DNA.
In this Colloquium we review several physical approaches to DNA detection
that have the potential to deliver fast and low-cost sequencing.
Center-fold to these approaches is the concept of nanochannels or nanopores which allow for
the spatial confinement of DNA molecules. In addition to
their possible impact in medicine and biology, the methods
offer ideal test beds to study
open scientific issues and challenges in the relatively unexplored area
at the interface between solids, liquids, and biomolecules at the nanometer
length scale. We emphasize the physics behind these methods and ideas, critically
describe their advantages and drawbacks, and discuss future research opportunities in this field.
\end{abstract}

\renewcommand{\thefootnote}{\fnsymbol{footnote}}

%\date{February 2006}
\maketitle
\tableofcontents

\section{Introduction}
\label{sec:intro}

After the first sequencing of the full human genome
~\cite{Venter2001-1,Lander2001-1} and decades of sequencing
improvements ~\cite{Chan2005-1}, genome-based
medicine has come ever closer to reality. A low-cost and rapid method to
sequence DNA would dramatically change the way we do medicine, and
would give us new tools to study biological functions and evolution.

This new technology would allow us to sequence and compare a plethora of individual human genomes,
enabling us to locate sequences which cause hereditary diseases and to discover mutated sequences.
Individual medical consumers can then be tested for these known genetic defects. Thus, medicine and treatments
can be tailored to their specific condition(s). This latter goal is known as {\it personalized
medicine} and would be a tremendous advancement in the medical field.
Furthermore, the genomic information gathered could be used in the study of biology by, e.g., comparing
genomes between species to find common properties and functions. This information could be used both medically (e.g., in
animal testing of potential cures for humans) and also to examine the evolutionary heritage of all species.

In order to reach the goal of a rapid and low-cost
sequencing method, one cannot rely only on current techniques, which
involve costs of about 10 million USD and several months time to
sequence a single human genome~\cite{Fredlake2006-1}. Improvements
of current technology face both fundamental and practical limitations, such
as a small -- on the order of a 1000 bases -- read-length limit using
electrophoresis~\cite{Chan2005-1}, which will restrict the impact of
subsequent developments in this direction. Therefore, radically novel ideas
need to be implemented and demonstrated to be
cost-effective and accurate.

There are several candidates which may fill this role~\cite{Deamer2000-1,Deamer2002-1,
Akeson1999-1,Kasianowicz1996-1,Kasianowicz2001-1,Zwolak2005-1,Lagerqvist2006-1,Gracheva2006-1,Heng2005-1,Lindsay2006-1,NHGRI,Lee2007-1}.
All these candidates have one trait in common: they employ nanoscale probes to examine the structural
or electronic signatures of individual DNA bases. That is, they rely on physical differences between the bases. This
is a major departure from existing sequencing paradigms which rely on chemical techniques and physical differences of strands
of DNA.
%\footnote{We differentiate between the terms
%``chemical'' and ``physical'' based on the use of a chemical process.
%If, for instance, a species is added to the DNA that changes its atomic makeup
%or its atomic structure, then a chemical process has occurred. Current techniques
%use a chemical process and then detect physical
%differences (in the form of optical emission) of the atomically modified strands.
%In principle, a chemical process can enhance the detection capabilities
%of the nanoscale physical detection schemes.
%However, they would add a potentially slow extra step to the full sequencing procedure. Nonetheless,
%detection of modified bases -- and this  could mean a  base completely replaced with some other species -- may be
%significantly easier. It is our hope that this Colloquium will set the stage for such developments by making it clear what physical
%signals are detectable and how different the signals are for the individual bases.}.

Most importantly, these proposals challenge our understanding of,
and ability to manipulate and probe, physical processes at the
interface between solids, liquids, and biomolecules down to the
nanometer scale regime~\cite{Diventra2004-2}. Indeed, in order
to understand the feasibility, speed, and accuracy of these novel
approaches we are naturally led to examine several physical
questions about the individual bases and the influence of the
solid/liquid environment:
\begin{itemize}
\item What is the difference in magnitude of physically measurable properties between the bases?
\item How do the nucleotide structural dynamics affect the measurable signals?
\item How do the different bases interact with the components of
the detection apparatus, e.g., the nanopore, the surfaces, the
electrodes, a scanning probe tip, or the other molecules present?
\item How does the atomic makeup and structure of the different bases affect the surrounding
fluid and ionic dynamics? And vice versa, how do the latter affect the structure and electronics of the bases?
\item What are the significant sources of noise?
\end{itemize}

In addition, many of the suggested sequencing methods rely
on nanopores either as a housing to contain the nanoscale probe(s) or as a restriction that causes differentiation in
some signal between the bases. Thus, the fabrication of the nanopores and the DNA translocation dynamics have an important
bearing on the following questions:
\begin{itemize}
\item How well can one probe the DNA on the single base scale according to the
dimensions and ``uniformity'' of nanopores?
\item How fast can one probe base differences?
\item What are the limits on read-lengths?
\item How fast and regular does the DNA translocate through the pore?
\end{itemize}

The above points beget even more basic and general physical questions:
\begin{itemize}
\item What is the meaning and role of electronic screening at the nanometer scale and in a
strongly confined and fluctuating environment?
\item What is the meaning of capacitance, thermal energy, charging energy, etc.,
under these atypical conditions and at small length
scales? How do these quantities evolve into their respective bulk properties?
\item Do liquids show unexpected dynamical features at the nanoscale?
\item How do electrons move in ``soft'' materials and dynamical environments?
\end{itemize}

These questions will accompany us for the full length of this Colloquium.
We stress their importance for the detection and sequencing approaches, review some
partial answers found in the existing literature, and point out possible future research directions to explore them in more depth.

The Colloquium is organized as follows: In Sec.~\ref{sec:curr} we give a
very brief account of current sequencing techniques. This primer will help
the reader get familiar with the state of the art in this field.
In Sec.~\ref{sec:phys} we outline the physical properties of DNA and its bases.
%This can be divided into two parts, its
%structural characteristics, including secondary structure and atomic makeup (Sec.~\ref{sec:struct}), and
%its electronic structure (Sec.~\ref{sec:elect}).
In Sec.~\ref{sec:nano}, we discuss nanopores as a useful
building block for rapid DNA sequencing and detection.
%We first give the basic concepts in Sec.~\ref{sec:concept},
%then we detail experimental results and accomplishments in Sec.~\ref{sec:exper}, translocation dynamics of DNA
%in Sec.~\ref{sec:trans}, and pore electronics in Sec.~\ref{sec:ion}.
After this, we move on
to physical approaches to DNA sequencing in Sec.~\ref{sec:seq}.
%We divide this Sec. into three parts depending
%on the type of signal measured: electronic signal detection in ~\ref{sec:seqelect}, optical signal detection in ~\ref{sec:seqoptic},
%and force detection in ~\ref{sec:seqforce}.
Finally, we conclude in Sec.~\ref{sec:conclu}.

\section{Current sequencing techniques}
\label{sec:curr}

\setcounter{footnote}{0}

A thorough introduction to existing sequencing methods can be found in~\onlinecite{Chan2005-1},
~\onlinecite{Fredlake2006-1},~\onlinecite{Dovichi2000-1}, and ~\onlinecite{JGI};
references therein provide a more technical account.
Present day sequencing methods are an improved version of the Sanger
method~\cite{Sanger1977-1}. The sequencing
process can be divided into four overall steps~\cite{Chan2005-1}:
(1) DNA isolation, (2) sample preparation, (3) sequence production,
and (4) assembly and analysis. Step (1) is simply the isolation of
the strand of DNA which needs to be sequenced. Within step (2) the
DNA needs to be replicated and also broken into (many) very short
strands. The length of the strands is dictated by the actual
sequencing technology used.

\begin{figure}
\begin{center}
\includegraphics*[width=5.5cm]{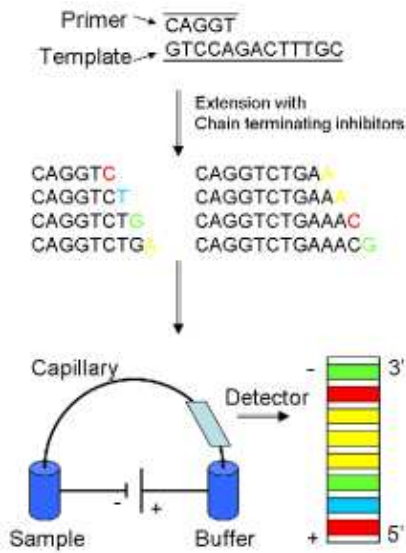}
\caption{(Color in online edition) The three processes used to produce the
sequence of a strand of DNA. One starts with a DNA strand to be
sequenced, called the template, to which one adds a primer which is
complementary to part of the template. A DNA polymerase reaction extends
the sequence starting from the primer by adding nucleotides
to the $3^\prime$ end. In the presence of a smaller
amount of dideoxynucleotides, the chains terminate at
various places along the complementary strand, and after denaturation,
produce single strands of
different length. These strands are then sorted with capillary
electrophoresis and detected by laser excitation of fluorescent tags
(one can either use multiple lanes each with its own
dideoxynucleotide and fluorescently tagged primer, or use
fluorescently tagged dideoxynucleotides). The detection information
is then sent to a computer for assembly and post-processing. The
terminology within this caption is defined in
Sec.~\ref{sec:struct}.} \label{current}
\end{center}
\end{figure}

Step (3) combines three components for the detection of the bases
in the DNA sequence, as shown in Figure~\ref{current}. {\em First}, chemical
elongation creates labeled strands of DNA with the random insertion of a
chain-terminating nucleotide (introduced by~\onlinecite{Sanger1977-1}).
{\em Second}, an electrophoretic\footnote{``Electrophoresis'' is a general term meaning the
action of driving charged molecules/particles in a solution with an electric field. More specifically,
it is taken to mean driving the particles through a porous matrix.}
process spatially separates the different lengths of DNA in a porous matrix. {\em Third}, an optical readout
detects the fluorescent end groups or primers, which indicates the last
base on each of the different lengths of DNA.

Step (4) is the post-processing of the sequence data, which 
involves the reassembly of the short sequences to get the complete sequence of the
original strand of DNA. Because of this, the short strands of DNA are required to have
large overlapping sequences in order to match them up.

From above, one sees that the current methods
rely on very complex sample preparation and post-processing
of the data. One of the main causes of this complexity
is a fundamental barrier to the maximum {\it read-length} achievable
when using electrophoresis~\cite{Chan2005-1}.\footnote{Practical application of
the chain-terminating chemistry can also limit read-lengths.}
The read-length is the longest strand of DNA which can be sequenced
accurately and efficiently within step (3).
The read-length is limited because electrophoresis is sensitive to
the physical difference between different lengths of (single-stranded) DNA. Thus, intuitively, one
expects that as the strand gets longer, distinguishing a strand of $N$ bases with
 one of $N+1$ bases becomes increasingly
difficult because the percent difference in the strand properties tends to zero.
As a consequence, if a sequencing scheme
allows for longer read-lengths of the DNA, this simplifies sample preparation and
post-processing. Since the physical schemes described in this review rely on single nucleotide
detection, the amount of sample preparation should be considerably reduced.
The electrophoresis step is also intrinsically slow~\cite{Chan2005-1}.

This review focuses on methods which improve/modify step (3) of the sequencing process.
However, as we just saw in the case of the read-length,
the four sequencing steps are not independent. Modifications to step (3) can
reduce the time and complexity of steps (2) and (4) and correspondingly reduce their costs. Thus, the technological
motivation for improving step (3) comes
from its pivotal role in the sequencing process.

\section{Physical characteristics of DNA}
\label{sec:phys}

\setcounter{footnote}{0}

In this section we give an introduction to the basic properties and
structure of the polynucleotides (PN)\footnote{One
also refers commonly to {\em oligonucleotides}, indicating a short strand of DNA/RNA.}
and their constituents. The bases are
very similar to each other and, thus, these properties are crucial
to understanding their distinguishability via size, electronic
states, or interactions with the surroundings.

\subsection{Structure}
\label{sec:struct}

\begin{figure}
\begin{center}
\includegraphics*[width=6.5cm]{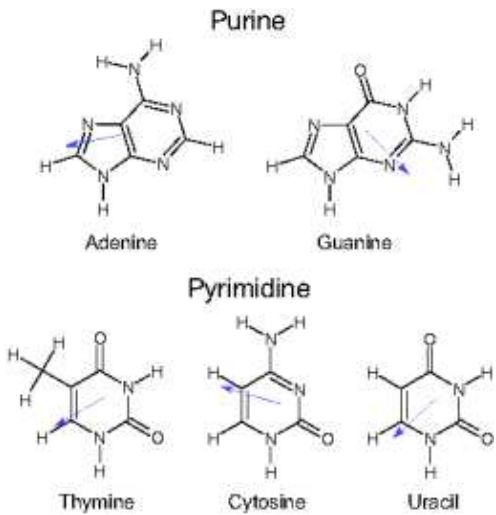}
\caption{(Color in online edition) Atomistic structure of the
nitrogenous bases found in DNA and RNA.
For DNA, these are the purine bases adenine (A) and guanine (G), and the pyrimidine bases
thymine (T) and cytosine (C). For most ribonucleic acids, uracil (U) takes the place of
thymine. Each base
is shown so that the bottom-most hydrogen indicates where the base attaches to the
sugar group of the backbone. Also shown are the directions of their dipole
moments.}
\label{bases}
\end{center}
\end{figure}

Both deoxyribonucleic acid (DNA) and ribonucleic acid (RNA) are built up of different
bases attached to a sugar-phosphate backbone.
The five bases that make up these polymers are
shown in Fig.~\ref{bases}. These can be classified
into two categories: the {\it purine} bases (A and G) and the
{\it pyrimidine} bases (C, T, and U). The purine bases consist of a six- and a five-membered
ring with a common edge. The pyrimidines have just a six-membered ring.
These classifications are natural based on the chemical structure.
We will see that for many physical properties the distinction into
purines and pyrimidines is not helpful.
However, one would expect that some physical properties would divide along this
classification. For instance, purines are larger and
thus one might expect that they would interact more
strongly with surfaces in a confined space.

\begin{figure}
\begin{center}
\includegraphics*[width=6.5cm]{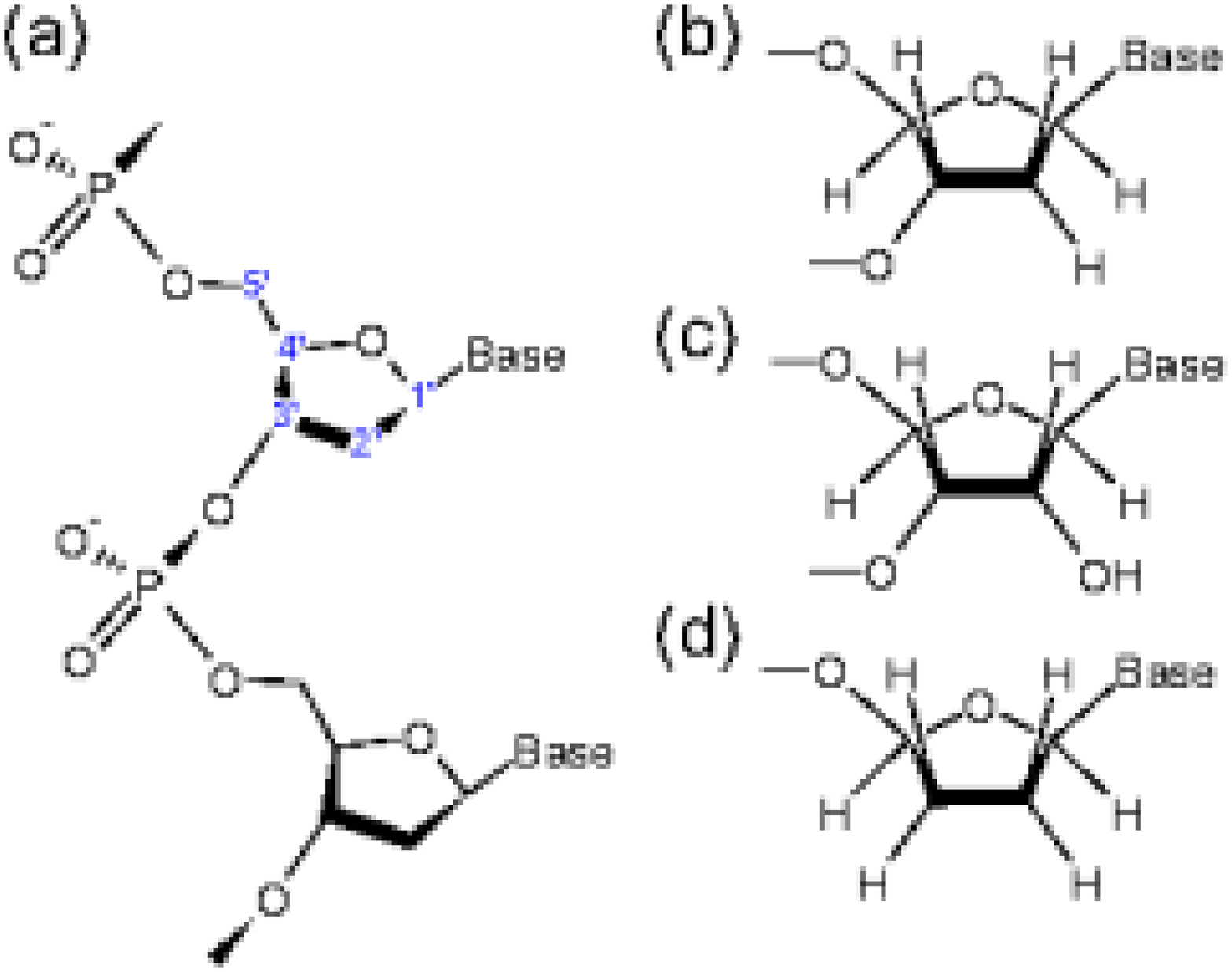}
\caption{(Color in online edition) The atomistic structure of the nucleic acid backbones. (a) Two repeat units
of the deoxyribonucleic acid backbone.
The phosphate groups have a negative charge
in solution, which is shared between the two oxygens solely bound to the phosphorous atom.
Also shown is the numbering system. The sugar carbon attached to the base
is labeled $1^\prime$ (the prime indicating that the numbering is on the
sugar), then the other carbons are labeled around the ring up to $5^\prime$. This
is the origin of the terminology ``$3^\prime$'', ``$5^\prime$''. A single
strand can end after the
$3^\prime$ carbon or the $5^\prime$ carbon.
The rest of the figure gives the atomistic structure, as a unit in a polynucleotide,
of the (b) deoxyribose sugar, (c) ribose sugar,
and (d) dideoxyribose sugar. The latter is used in the Sanger method, see
Sec.~\ref{sec:curr}, because the absence of the 3$^\prime$ hydroxyl group
will not allow further chain extension. The typical nomenclature is to add a ``d'' in front of
the base to indicate that it is attached to a deoxyribose sugar instead of a ribose sugar,
e.g., poly(dA) is a polynucleotide of DNA, while poly(A) is a polynucleotide of RNA.}
%The ``deoxy'' just means an ``oxy'' (hydroxyl group) has been removed.}
\label{backbone}
\end{center}
\end{figure}

The backbone structure of a polynucleotide is shown in
Fig.~\ref{backbone}a. Each monomer unit is a {\em
nucleotide}, which consists of a base, phosphate group ($\mathrm{PO}_4$), and sugar. 
For DNA (RNA), the latter is the deoxyribose (ribose) sugar
shown in Fig.~\ref{backbone}b(c). The difference of only a single
hydroxyl group ($\mathrm{OH}$) between RNA and DNA can be important
for the global structure of the polynucleotide. Within the repeat
unit of the polynucleotides there is one phosphate group, i.e., a nucleotide monophosphate. However,
triphosphate monomers are used in chain extension reactions.
The formation of the double-stranded (ds) DNA helix
 occurs by the Watson-Crick base pairing, where
A pairs with T, and G pairs with C.

One of the most important properties of polynucleotides is that they
are charged in solution. The pK$_a$ of the phosphate group, i.e.,
the measure of how readily that group will give up a hydrogen cation
(proton), is near one. Thus, under most ionic conditions (including
physiological pH), the backbone will contain a single negative
charge for each nucleotide unit (or two negative charges for a
Watson-Crick pair of nucleotides in a double strand). In solution,
though, there will be, on average, nearby counterions such as sodium
(Na$^+$), potassium (K$^+$), or magnesium (Mg$^{2+}$), which
neutralize a part of this charge. The fact that the strand of
nucleotides is charged is what allows one to pull the DNA through
nanopores with an electric field.

There are other important properties that will help us understand
the experiments and theoretical proposals below. For instance, a
polynucleotide has a global orientation, with one end
a 5$^\prime$ and the other end a 3$^\prime$, as described in
Fig.~\ref{backbone}. We will see in Sec.~\ref{sec:nano} that this
is important for the structural dynamics of polynucleotides translocating
through a nanopore.

There is also a global property called {\em secondary structure}.
For ds-DNA, for instance, there exist several different types of
helices. The most common ones are called A-DNA and B-DNA. What
changes the global structure of DNA between A- and B-DNA is the
ionic and water environment. B-DNA is preferred in aqueous
environment because water molecules can bind in the groves along the
helix.
 The base pair-base pair distance
in the B-DNA helix is 3.4 $\mbox{\AA} \;$ and there is a 36
degree angle between them, which gives about 10 bases per turn of
the helix. The diameter of B-DNA is $\sim 2$ nm.
A-DNA has different physical dimensions. The difference between B-DNA and A-DNA is probably not
important for sequencing, but some proposals for detection of ds-DNA
will be affected by such a change in global structure due to
the environmental conditions. Furthermore, there is a
process called {\it denaturation} where the two strands in ds-DNA
unbind into single-stranded (ss) DNA molecules.

\begin{figure}
\begin{center}
\includegraphics*[]{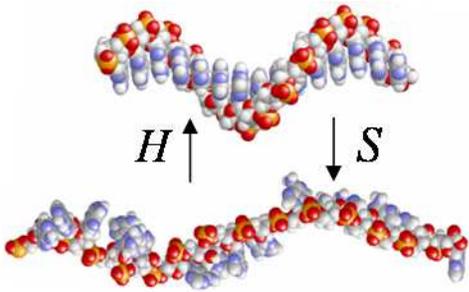}
\caption{(Color in online edition) Schematic of a secondary structure of
ss-RNA. On the top, stacked bases forming a
single-stranded helix due to enthalpic effects (denoted as $H$ in
the figure). Entropic effects (denoted with $S$ in the figure)
transform this structure into the one on the bottom, a random coil
polymer. The homogeneity of the strand is important for secondary
structure, as are the ionic conditions and the temperature. For
instance, poly(C) at room temperature and neutral pH is mostly a
single-stranded helix 1.3 nm in
diameter~\cite{Arnott1976-1}.}
\label{secstruct}
\end{center}
\end{figure}

Single strands show secondary structure as well, which also
depends on ionic conditions and temperature. A schematic of
secondary structure in ss-RNA is shown in Fig.~\ref{secstruct}.
Generally, one can think of secondary structure as a result of the competition
between enthalpic and entropic factors. The interaction energy of
base stacking is the main factor favoring secondary
structure~\cite{Searle1993-1},
such as the helix in Fig.~\ref{secstruct}. On the other hand, the entropic
factors are mainly due to rotations of backbone degrees of
freedom~\cite{Searle1993-1}. If one raises the temperature, entropic
factors will dominate and bring the strand to a {\em random coil}
form. When this transition happens depends on the bases in the
strand~\cite{Vesnaver1991-1}, ionic conditions~\cite{Dewey1979-1},
and other environmental factors~\cite{Freier1981-1}. The effect of secondary
structure has already been seen in the ionic currents through
nanopores~\cite{Akeson1999-1,Meller2000-1}.

Another type of secondary structure is a {\em hairpin}. RNA/DNA
hairpins are single strands of RNA/DNA that wrap around to form a
double strand. However, in doing so, they have to form a {\em loop} of
unpaired bases, which causes an unfavorable strain on their formation.
The double-stranded portion is called the {\em stem}, and
sometimes a single-stranded portion can exist on the chain as well.
This latter type of hairpin is used in some of the experiments described
below.

Let us also add that in some of the proposals for sequencing, an exonuclease could be
used to chop the DNA up into individual nucleotides.
A nuclease is an enzyme that acts as a catalyst to polynucleotide breakup,
e.g., the linking oxygen between nucleotide repeat units can be hydrolyzed
to two OH groups. An {\em exonuclease} is a nuclease that ``eats away''
at the ends of the polynucleotide, breaking off one nucleotide at a time.

Some particular properties of interest to us are the base and
nucleotide sizes. The sizes of the DNA bases and corresponding
deoxyribonucleic acids are shown in Table~\ref{sizes}. Using
geometries from~\onlinecite{Zwolak2005-1}, these sizes were computed
by drawing a sphere around each atom of its van der Waals radius,
and taking the union of these spheres. From the table, one can
observe that even in the most ideal case, the size of the base alone
is unlikely to provide a distinguishable signal. This will be
discussed at length in Sec.~\ref{sec:ion}.

\begin{table}
\caption{Sizes of the DNA bases ($V_B$) and nucleotides ($V_N$) in $\mbox{\AA}^3$.
Also given are the surface areas $A_B$ and $A_N$ in $\mbox{\AA}^2$.
We define the fraction of free volume left, $F_d = (V_p-V_N)/V_p$,
where the pore volume, $V_p=\pi l d^2/4$, is for a cylinder of length $l$ and diameter $d$.
We assume each nucleotide occupies a length $l \approx 7 \mbox{\AA}$, which is its own length in a
completely extended strand.
We consider pore diameters of 15 and 20 $\mbox{\AA}$ which are approximately
equal to the $\alpha$-hemolysin pore
diameter in its stem. For comparison, the backbone and U
volume (surface area) are 214 (225) and 128 (142), respectively.}
\label{sizes}
\begin{tabular*}{8.5cm}{ c @{\extracolsep\fill} c @{\extracolsep\fill} c @{\extracolsep\fill} c }
Base & $V_B$ ($A_B$) & $V_N$ ($A_N$) & $F_{15}$ ($F_{20}$) \\
\hline \hline
A & 157 (166) & 349 (340) & 0.72 (0.841) \\
G & 168 (177) & 359 (351) & 0.71 (0.837) \\
C & 133 (147) & 324 (319) & 0.74 (0.853) \\
T & 150 (163) & 339 (331) & 0.73 (0.846) \\
\hline \hline
\end{tabular*}
\end{table}

\subsection{Electronics}
\label{sec:elect}

Two detection/sequencing schemes
propose to use transport and voltage fluctuations to
distinguish the bases. These will be sensitive to the
electronic structure of the different nucleotides and also
to the geometry and local environment.

\begin{figure}
\begin{center}
\includegraphics*[width=7.5cm]{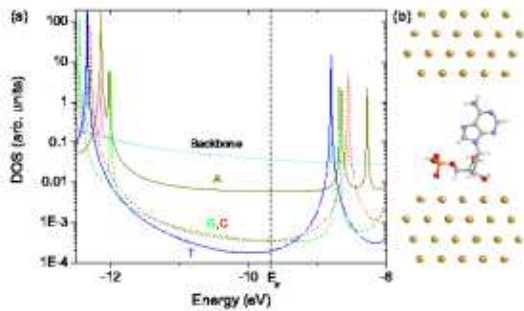}
\caption{(Color in online edition) (a) Projected density of states
of the DNA nucleotides between two gold electrodes shown in (b), as
computed in~\onlinecite{Zwolak2005-1} with tight-binding parameters.
$E_F$ is the Fermi level of gold within the same approach.}
\label{DOS}
\end{center}
\end{figure}

Transport through nucleotides will detect differences in electronic
states via their energy and spatial extension. To understand the
differences between the bases we compute the density of states of
the nucleotides (placed between two electrodes) projected onto the
backbone and bases, as shown in Fig.~\ref{DOS}. The plot shows that
the different nucleotides do not have a considerably different
electronic structure, having molecular states very near in energy
compared to the position of the Fermi level. Most of the density of
states around the Fermi level is contributed by the backbone. This
is in part due to the upright geometry in Fig.~\ref{DOS}(b), which
allows for a large contact area of the backbone with an electrode.
The different spatial extension of the bases causes their
contribution to the DOS at the Fermi level to differ. These
differences will be heavily influenced by the nucleotide geometry
and orientation.

\begin{figure}
\begin{center}
\includegraphics*[width=7.5cm]{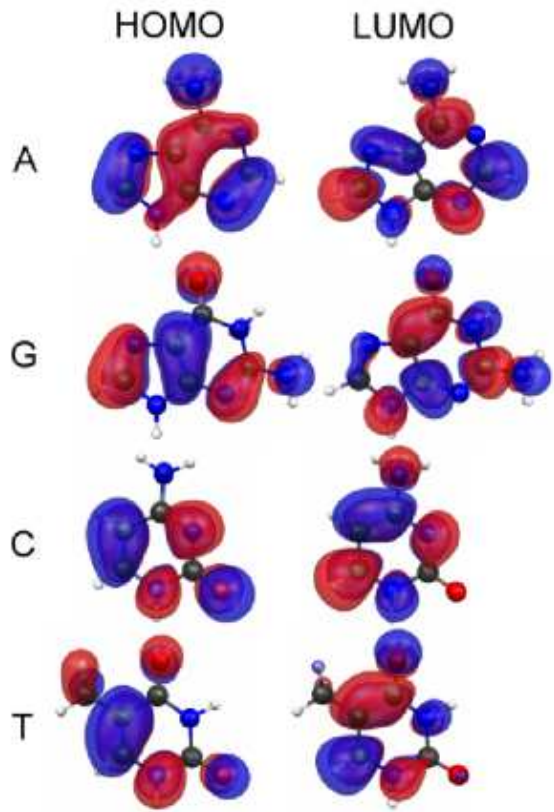}
\caption{(Color in online edition) Isosurfaces of the HOMO and LUMO states
of the isolated bases.
The red and blue colors indicate opposite signs of the wavefunction.
The HOMO and LUMO states do not change from the
individual base to the passivated nucleotide. However, when the
backbone becomes charged in solution, the HOMO and LUMO states will
shift in both character and energy.}
\label{states}
\end{center}
\end{figure}

However, the basic input to determine the base-electrode coupling is
the character of the molecular states. For the isolated bases, this
is shown in Fig.~\ref{states} for the highest-occupied molecular
orbital (HOMO) and lowest-unoccupied molecular orbital (LUMO). For
nucleotides only weakly contacted with electrodes, the character of
these states is going to determine how well the nucleotide can
couple to charge carriers in the electrodes. We can see that for all
the bases, the states are distributed around the rings. These states
remain that way even in the presence of the passivated or charged
backbone (although they may not remain the HOMO and LUMO states of
the nucleotide). The spatial extension of the base is roughly
correlated with the relevant electronic states and, thus, will
determine a large part of the coupling of the nucleotide to the
electrodes.

On the other hand, other electronic approaches may measure the
dipole and higher moments. The dipole moment magnitudes are shown in
Table~\ref{dipoles}, with the corresponding directions shown in
Fig.~\ref{bases}. The dipole moments are for the isolated DNA bases
and for the corresponding nucleotides. The moments and the molecular
wavefunctions were computed within Hartree-Fock using the geometries
in~\onlinecite{Zwolak2005-1}.

\begin{table}
\caption{Theoretically calculated dipole moments of the bases
($p_B$) and the deoxyribonucleic acids ($p_N$) in Debye ($\sim 0.21
e \, \mbox{\AA}$). For comparison, the dipole moment of U is 4.79 D
($p_U^{exp}=4.2$ D) and the backbone dipole by itself is 1.97 D.
Experimental values, $p_B^{exp}$, are from
~\onlinecite{Kulakows1974-1},~\onlinecite{Weber1990-1},
and~\onlinecite{Devoe1962-1}.} \label{dipoles}
\begin{tabular*}{8.5cm}{c @{\extracolsep\fill} c @{\extracolsep\fill} c @{\extracolsep\fill} c }
Base & $p_B$ & $p_B^{exp}$ & $p_N$ \\
\hline \hline
A & 2.33 & 2.5 & 4.76 \\
G & 7.17 & 7.1 & 7.76 \\
C & 7.22 & 7.0 & 8.55 \\
T & 4.72 & 4.1 & 7.56 \\
\hline \hline
\end{tabular*}
\end{table}

Finally, a very important property of PNs relevant to the studies
discussed in this review, relates to its bonding properties at
surfaces, and in particular at the interior surfaces of nanopores.
These interactions, and their effect on the PN dynamics in confined
geometries, is a property of both the nucleotides and the type of
surface, thus making it a very complex issue. We defer its
discussion to later sections on particular experiments.

\section{Nanopores and polynucleotides}
\label{sec:nano}

\setcounter{footnote}{0}

About a decade ago, Kasianowicz and collaborators were able to pull
single-stranded polynucleotides through a biological nanopore by
applying a voltage across the pore which pulls on the charged PN
backbone~\cite{Kasianowicz1996-1}. These authors have detected the
translocation of the PN via measurement of the {\it blockade
current}, to be discussed below. Since this pioneering experiment,
nanopores have been used to extract a variety of information
characterizing the translocation, dynamics, and interactions of both
single- and double-stranded PNs in
nanopores~\cite{Kasianowicz1996-1,Kasianowicz2001-1,Henrickson2000-1,Akeson1999-1,
Meller2000-1,Meller2001-1,Butler2006-1,Chang2004-1,Chen2004-1,Deamer2002-1,Fologea2005-1,
Mathe2005-1,Storm2005-1}. Further progress is being made by moving
to synthetic nanopores where techniques are being developed to
control nanopore shapes, sizes, and other
characteristics~\cite{Li2001-1,Li2003-1,Storm2003-1,Wanunu2007-1}. Successful
sequencing and detection will require control of these
characteristics.

In addition to experiments, theoretical and computational work is
underway to help understand polynucleotide translocation through
nanoscale pores. There are two fruitful approaches to this complex
issue: phenomenological models and molecular dynamics. The
phenomenological models provide a highly reduced description of the
polymer dynamics, but they are able to elucidate the dependence of
the dynamics on parameters such as the polymer length, pore
dimensions, and applied field~\cite{Lubensky1999-1,
Sung1996-1,Ambjornsson2002-1,Loebl2003-1,Chuang2002-1,Kong2002-1,Chern2001-1,Muthukumar1999-1,
Slonkina2003-1,Meller2003-1,Muthukumar2001-1,Luo2006-1,Matysiak2006-1}. On the
other hand, if one wants to understand how to probe physical
differences between the bases, an atomistic description of the
polynucleotide dynamics is necessary. Molecular dynamics simulations
coupled to other computational methods have been used in this
context to study the signals and fluctuations expected when
measuring different physical quantities as the PN translocates
through the
pore~\cite{Lagerqvist2006-1,Lagerqvist2006-2,Lagerqvist2006-3,
Jenkins2005-1,Heng2005-1,Heng2005-2,Heng2006-1,Heng2004-1,
Aksimentiev2004-1,Gracheva2006-1,Gracheva2006-2,Cui2004-1,Muthukumar2006-1}.

In this section, we first review the basic concepts of the nanopore-polynucleotide experiment.
We will then move on to discuss some experimental results on biological and synthetic pores.
We also include a number of interesting physical results on PN translocation and
pore electronics, which demonstrate the wealth of fundamental physics that is contained in this new field.

\subsection{Nanopore-Polynucleotide concept}
\label{sec:concept}

\begin{figure}
\begin{center}
\includegraphics*[width=5.5cm]{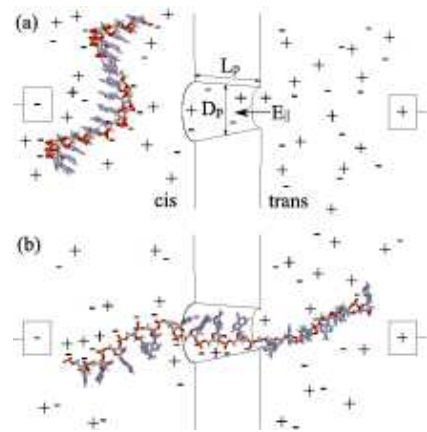}
\caption{(Color in online edition) Schematic of the nanopore-PN experiment. A membrane divides the
solution into what is commonly called the {\em cis} and {\em trans} chambers.
A bias across the separation creates a field $E_{\parallel}$ across
the membrane and drives
an ionic current through the pore as shown in (a). The field also pulls on the
negatively charged PN backbone, which causes the PN to get captured by and then
translocate through the
pore as shown in (b). While the PN is within the pore, ions are partially prevented from
occupying and flowing through the pore, thus reducing the ionic current. The pore
is characterized by its length $L_p$ and average diameter $D_p$.}
\label{poredim}
\end{center}
\end{figure}

\begin{figure}
\begin{center}
\includegraphics*[width=5.5cm]{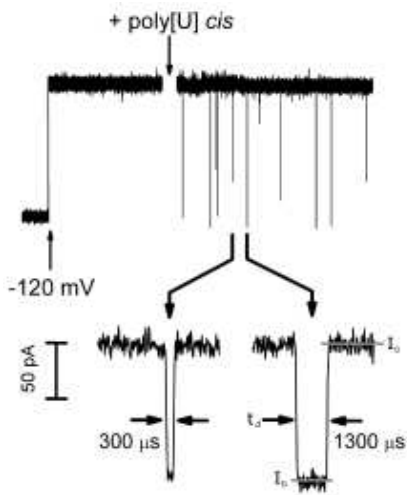}
\caption{Example of an ionic current signal. Upper Figure: A voltage of 120 mV
drives a current through a $\alpha$-hemolysin pore. When poly(U) is added to the cis chamber, it generates
blockade events. Lower Figure: Two examples of blockade events.
The average open pore current, $I_o$, the average blockade current, $I_b$,
and the translocation
duration, $t_d$, are all indicated on the figure.
Adapted from~\onlinecite{Kasianowicz1996-1}.}
%copyright US National Academy of Sciences
\label{blockade}
\end{center}
\end{figure}

Prior to \onlinecite{Kasianowicz1996-1}, researchers were already
interested in the ability to pull small charged molecules and
polymers through ion
channels~\cite{Bezrukov1993-1,Kasianowicz1995-1,Henry1989-1,Bustamante1995-1,Bezrukov1996-1,Bayley1994-1,Bezrukov1994-1}.
%
%Other references that look at proteins:
%Brundage, L., Hendrick, J. P., Schiebel, E., Driessen, A. J. M. &
%Wickner, W. (1990) Cell 62, 649–657.
%Akimaru, J., Matsuyama, S. I., Tokuda, H. & Mizushima, S.
%(1991) Proc. Natl. Acad. Sci. USA 88, 6545–6549.
%Goerlich, D. & Rapoport, T. A. (1993) Cell 75, 615–630.
%Simon, S. M. & Blobel, G. (1991) Cell 65, 371–380.
%
Thus, the basic question was whether PNs could also translocate
through, and be detected by, a pore. A schematic of this process is
shown in Figure~\ref{poredim}. The figure shows a membrane/thin
layer which divides a solution into two halves. An electrode is
placed into each halve. Without the PN, the electrodes pull the ions
through the channel to create an ionic current $I_o$ of the open
channel (see Figure~\ref{blockade}).

Due to the negative charges on the phosphate groups, the PN is pulled to the
positively biased (cathode) half of the solution.
Eventually the strand is {\em captured} and enters the pore. During a time
$t_d$, the {\em translocation duration}, the strand partially blocks ions from
the pore as shown in Figure~\ref{poredim}(b).
Even though the nucleotide is charged, it carries very little ionic current
through the pore because of its slow velocity compared to the ions.
Thus, its presence gives the blockade current
$I_b$ shown in Figure~\ref{blockade}, so that the PN translocation event
can be detected.
The translocation events also provide a way to
obtain information about the PN, such as its length and limited information on
composition and dynamics.

The actual values of the ionic current of the open pore and the
blockade current are due to several factors. The open pore
resistance is determined by both an energy penalty (mainly electrostatic) and
an entropic barrier to bring charges into the pore. We will see
below in subsection~\ref{sec:ion} that this is quite a complicated
problem in itself, due to the microscopic details of electric fields
in reduced geometries and fluctuating environments (e.g., the value
of a screened charge in an electric field at the nanoscopic level
can not be simply viewed as a charge within a dielectric
environment), and also the surface physics of pores in solution. The
blockade current is thus also a complex phenomenon. We will discuss
this issue in the context of the particular experiments described
below.

\subsection{Nanopore characteristics}
\label{sec:exper}

The current state of the art is divided into two lines of research.
One line uses biological pores, generally $\alpha-$hemolysin
pores~\cite{Song1996-1}.
%\footnote{There are other biological pores which DNA can
%translocate, see, for instance,~\onlinecite{Szabo1998-1} and
%~\onlinecite{Szabo1997-1}.}.
These pores have the right size scale
to detect differences in strands of PN. However, tailoring
characteristics, such as the pore diameter, is not straightforward.
The other line of
research operates with synthetic pores, which can be reasonably
controlled down to the sub-nanometer range.
One can also imagine making hybrid devices to help control pore properties.

The characteristics of the nanopore and the ability to control them
will be extremely important for sequencing methods. For instance,
methods which propose the use of nanoscale probes embedded in the
pore will require a pore size that maximizes the signal difference
between the bases beyond the many unavoidable sources of noise. In
most cases this means a pore diameter with the same width as a
single strand of DNA. However, the maximization of the signal
difference has to be balanced with other effects such as DNA-surface
interaction, which is minimal with a large diameter pore, and DNA
capture/translocation, which will not occur at small pore sizes. In
addition, the pore has to be fabricated in a way to make possible
the embedding of a nanoscale probe. This will put restrictions on
the types of material for the pore and its sizes/shapes.

\subsubsection{Biological pores}
\label{sec:biopore}

\begin{figure}
\begin{center}
\includegraphics*[width=5.5cm]{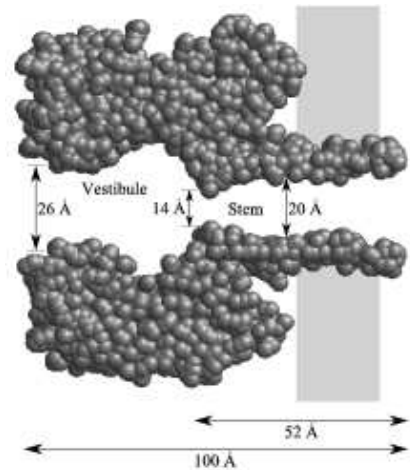}
\caption{Slab view of the $\alpha-$hemolysin pore. The uniform light
gray area represents the lipid bilayer into which the
$\alpha-$hemolysin assembles to make the pore. The initial part of
the pore with a much wider diameter is the {\em vestibule}. The
approximately 5 nm long neck is the pore {\em stem}. Some other
characteristics of the pore are a $\sim$18 nm$^3$ stem volume~\cite{Deamer2002-1}, and at 120 mV,
the pore current is 120 pA, giving almost $10^9$ ions/s.}
\label{alphapore}
\end{center}
\end{figure}

The $\alpha-$hemolysin pore is shown in Figure~\ref{alphapore}, with
some characteristics given in the figure caption.
%\footnote{The
%coordinates for the pore were taken from the Protein Data Bank (PDB
%ID: 7AHL; \onlinecite{Song1996-1}). All the distances are
%approximate distances obtained from atom-atom distances.}
In this
biological pore the smallest restriction that the PN has to
translocate through is $\sim$1.4 nm. Thus, ds-DNA, at a
diameter of 2 nm, can not translocate through the pore, but ss-DNA
can~\cite{Kasianowicz1996-1}. The pore is also both small and long enough that the PN
has to be locally extended, which gives an entropic barrier for
transport due to unraveling of the polymer. We give an
overview of some $\alpha$-hemolysin experiments in Table~\ref{tbio}.

\begin{table*}
\caption{Overview of some $\alpha$-hemolysin experiments. $T$ represents the
temperature and $v_{DNA}$ is the DNA velocity through the pore. ``p(X)'' stands for
 a polymer of X, and $L$ is the length of the polymer in nucleotides (nt). The
designation $3^\prime \leftrightarrow 5^\prime$ indicates an experiment which looked at the translocation direction.}\label{tbio}
\begin{tabular*}{16cm}{c @{\extracolsep\fill} c @{\extracolsep\fill} c @{\extracolsep\fill} c }
Ref. & PN & $L$ (nt) & Result \\
\hline \hline
\onlinecite{Kasianowicz1996-1} & p(U), others & $\sim150-450$ & $t_d \propto L$ \\
\onlinecite{Akeson1999-1} & p(A),p(C),p(dC),p(U),p(A)(C) & $\sim100-200$ & discrimination \\
\onlinecite{Meller2000-1} & p(dA),p(dC),p(dA)(dC),p(dAdC),others & $100$ & $t_d \propto \frac{1}{T^2}$, discrimination \\ %p(dC)(dT),p(dCdT)
\onlinecite{Meller2001-1} & p(dA) & 4-100 & $I_b$ transition at $L\sim L_p$,$v_{DNA}$ \\
\onlinecite{Mathe2005-1} & p(A),hairpin & $\sim 50$ &  $3^\prime \leftrightarrow 5^\prime$ \\
\onlinecite{Butler2006-1} & p(A),p(C),p(A)(C),p(C)(A) & 50,75 & $3^\prime \leftrightarrow 5^\prime$, discrimination \\
\hline \hline
\end{tabular*}
\end{table*}

\begin{figure}
\begin{center}
\includegraphics*[width=5.5cm]{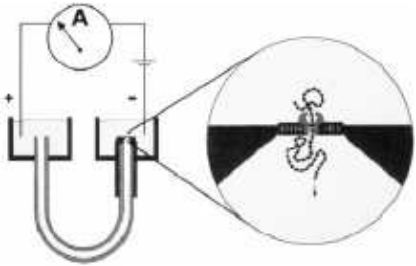}
\caption{A typical experimental setup for an $\alpha$-hemolysin pore.
The positive voltage is applied
to the trans chamber and the negative voltage to the cis one.
From~\onlinecite{Akeson1999-1}.}
%copyright Biophysical Society
\label{alphasetup}
\end{center}
\end{figure}

A typical experimental setup with an $\alpha-$hemolysin pore is shown in Figure~\ref{alphasetup}.
There are two chambers, the cis and the trans, with a buffered solution, e.g., of KCl. Between
the two is a Teflon partition with a small
orifice where the lipid bilayer is formed. When the hemolysin subunits are added
they spontaneously form into the pore. The formation of the pore is detected by the appearance of
an ionic current between the two chambers.
%At that point, one can stop the formation of additional
%pores by flushing the chamber of the hemolysin subunits.

\begin{figure}
\begin{center}
\includegraphics*[width=5.5cm]{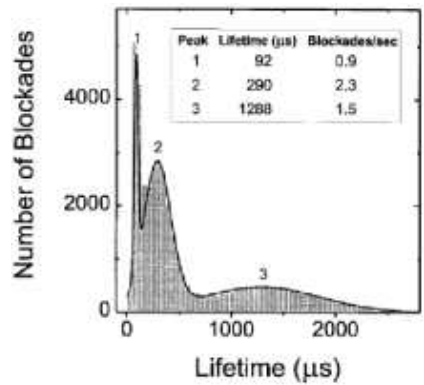}
\caption{Translocation duration (lifetime) of the ionic blockade
events measured by~\onlinecite{Kasianowicz1996-1}.
The events fall under three different lifetimes, see the text.
From~\onlinecite{Kasianowicz1996-1}.}
%copyright US National Academy of Sciences
\label{KasianowiczFig2}
\end{center}
\end{figure}

The experiment is then conducted by adding PNs into the cis chamber.
%\footnote{We describe some of the experiments that measure
%length dependence, composition, and global properties. Other
%experiments have examined the translocation under different
%conditions, such as temperature~\cite{Meller2000-1}, which give us
%additional information on PN dynamics. We refer the reader to some
%of the earlier reviews and introductory articles on biological
%pores~\cite{Deamer2002-1,Deamer2000-1,Meller2003-1}.}
The first experiments were done with
poly(U)~\cite{Kasianowicz1996-1}. Upon its addition, transient
blockades of ionic current were observed. The blockade events
actually fall under three different types distinguished by their
lifetime, as shown in Figure~\ref{KasianowiczFig2}. There are fast
blockades that are independent of the poly(U) length. Thus, most
likely, these are events where the strands just cover the entrance
to the pore or just partially enter the pore, but do not translocate
through it.
%I believe these fast events are now normally filtered out of the data from
%later experiments.
However, surprisingly, both the other two types of events have
lifetimes linearly dependent on the length of poly(U), and inversely
dependent on the applied voltage. ~\onlinecite{Kasianowicz1996-1} conjectured that this could be due to
different translocation speeds of strands entering with either the
$3^\prime$ or $5^\prime$ end. Later experiments confirmed this
conjecture by examining strands composed of two homogeneous blocks
and exploiting the pore's ability to distinguish between different
nucleotides, see below~\cite{Mathe2005-1}.

\subsubsection{Synthetic pores}
\label{sec:synpore}

Synthetic pores offer additional opportunities for detection and
sequencing of PNs. For instance, one can adjust the pore dimensions and
properties to meet the needs of a particular experiment.
Also, they open the possibility of integration of external sensors
and probes, such as transverse electrodes. In addition, the
parameter range for their operation is larger (although the $\alpha$-hemolysin
pore is quite robust, biological pores in general are
only open under certain voltages, ionic concentrations, and
temperatures~\cite{Hille2001-1}), and indeed different parameters
such as salt concentration, temperature, voltage, and viscosity may be key to
operating the pore in a regime where sequencing and detection is
possible or most optimal, because these conditions control
properties like the translocation velocity and capture rate (see, e.g.,~\onlinecite{Fologea2005-2,Henrickson2000-1}). We
give an overview of some of the synthetic pores made to date and the
translocation experiments in Table~\ref{tsyn}.

\begin{table*}
\caption{Overview of some synthetic nanopores. Next to each pore type, their diameters and lengths are
given as ($\{D_p\}$;$\{L_p\}$) in units of nanometers.}
\label{tsyn}
\begin{tabular*}{16cm}{c @{\extracolsep\fill} c @{\extracolsep\fill} c @{\extracolsep\fill} c @{\extracolsep\fill} c }
Ref. & Pore  & PN & $L$ (nt) & Result \\
\hline \hline
\onlinecite{Li2003-1} & Si$_3$N$_4$ (3,10;5-10) & bio-ds-DNA & 3,10 kbp & Folded DNA $I_b$ events \\
\onlinecite{Chen2004-1} & Si$_3$N$_4$ (15;n/a) & $\lambda$-DNA, other & 48.5, 10, 3 kbp & $v_{DNA} \propto V$, folded \\
\onlinecite{Storm2005-1} & SiO$_2$ (10;20) & bio-ds-DNA & $\sim$6-97 kbp & $t_d \propto L^{1.27}$ \\
\onlinecite{Fologea2005-1} & Si$_3$N$_4$ (4;5-10) & bio- ss- and ds-DNA & 3 kbp & denaturation \\
\onlinecite{Chang2004-1} & SiO$_2$ (4.4;50-60) & bio-ds-DNA & 200 bp & $I_b>I_o$ \\
\onlinecite{Heng2004-1} & Si$_3$N$_4$ (1,2.4;10,30) & p(dT),ds-DNA & 50-1500 bp & length discrimination\\
\hline \hline
\end{tabular*}
\end{table*}

The fabrication of synthetic nanopores is still in its nascent stages.
Mainly two groups have pioneered techniques for solid-state pore fabrication.
Golovchenko, working with J. Li, has developed a technique using
low-energy ion beam to sculpt (``ion-beam sculpting'') a nanoscale hole in
Si$_3$N$_4$~\cite{Li2001-1}. 
Dekker {\it et al}. have developed a technique based on an electron beam and a
visual feedback procedure~\cite{Storm2003-1}.

\begin{figure}
\begin{center}
\includegraphics*[width=6.5cm]{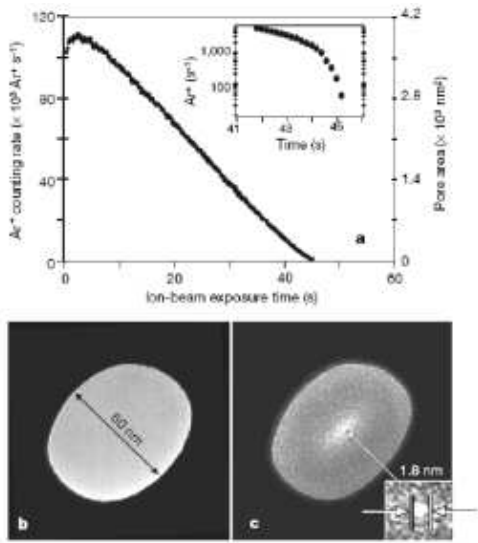}
\caption{Fabrication technique of~\onlinecite{Li2001-1} for creating nanoscale
solid-state pores: a Si$_3$N$_4$ membrane with a large pore is first created (panel b).
An ion beam is then focused at this large pore, activating a diffusion process which
closes the hole (panel c). The current of Ar$^+$ decreases as the pore shrinks (panel a),
the monitoring of which can be used to control pore size.
From~\onlinecite{Li2001-1}.}
%copyright Nature Publishing Group
\label{IonSculpt2}
\end{center}
\end{figure}

The technique developed by Golovchenko {\it et al}. is shown in
Figure~\ref{IonSculpt2}~\cite{Li2001-1}. One starts by creating a
large diameter pore in a solid-state membrane using a focused ion
beam (Figure~\ref{IonSculpt2}b). In their case, a $\sim 60$ nm pore
in a Si$_3$N$_4$ membrane is first created. Then this pore is
exposed to an Ar$^+$ beam, which, instead of knocking atoms off the
membrane and opening the pore further, activates a diffusion process
and the pore starts to close~\cite{Li2001-1}. The current of the
Ar$^+$ coming through the pore is directly dependent on the pore
area. Thus, one can measure the Ar$^+$ current, shown in
Figure~\ref{IonSculpt2}a, as a function of time, and use this to
controllably shrink the pore down to the nanometer scale
(Figure~\ref{IonSculpt2}c). The precise composition of silicon and
nitrogen around the pore is not known~\cite{Li2001-1}.

Dekker and co-workers have developed a similar technique based on
the idea of pore diameter reduction upon
irradiation~\cite{Storm2003-1}. They use a high-energy electron beam
and SiO$_2$ membrane. The initial, larger pore can be formed in
different ways.
%In one case, one chemically etches
% a relatively thick SiO$_2$ membrane to form a pore of
%dimensions in the tens of nanometers. In the other case, an
% electron beam is finely focused on very thin (10 nm) SiO$_2$ membranes. In both cases, after
But once the initial pore is made, a transmission electron microscope (TEM)
is used to shrink it. By using the imaging capability of
the TEM, one can visually watch the pore diameter
reduce~\cite{Storm2003-1}. Since the rate of size reduction is very
slow (0.3 nm per minute in these experiments), the visual images can
be used to monitor the diameter of the pore and the process stopped
when the desired size is reached. The resolution of the microscope
($\sim$0.2 nm) then sets the limit of accuracy for reaching the
desired size. However, due to the roughness of the surface, one can
only control the pore size to about 1 nm
dimensions~\cite{Storm2003-1}.

Synthetic pores have properties that
differ from the biological pores. The surfaces can be charged in
solution, creating an additional complication to understanding ionic
and PN transport. For instance, silicon oxide surfaces can have a negative surface charge
density on the order of $10^{-2}$ e/$\mbox{\AA}^2$ in aqueous solution~\cite{Chang2004-1}.
This would give quite a substantial charge within any reasonably sized nanopore, and
would have to be neutralized by counterions.
Biological pores, though, can
have unusual potential profiles which may have internal sites with
trapped charges~\cite{Hille2001-1}.

The observation that high energy beams can cause pores to
shrink is a very interesting phenomenon in and of itself, in addition to its
implications for the fabrication of nanoscale structures/pores.
We want to
mention here that~\onlinecite{Storm2003-1} have observed that
there is a transition from shrinking to growing as the pore diameter
is increased, and have explained this observation
by a surface tension effect.

\subsection{DNA translocation}
\label{sec:trans}

One can divide the translocation of PN through a pore into two categories: universal
properties of polymer dynamics (entropic forces, Brownian motion, charges and screening),
and specific properties that rely on the atomic compositions of the nucleotides
(e.g., interaction potentials with the pore surface). Depending on the quantity under
consideration, and the parameter regimes of the device, either one or both of these
categories will be important.

\subsubsection{Universal}

The two basic properties of charged polymer dynamics are the processes of capture
and translocation. The capture of the polymer will depend on the diffusion of the polymer
from the bulk to the pore and on local effects around the pore, such as the electric
field and interactions between the entrance of the pore and the
polymer. The capture rate will depend on concentration and applied bias~\cite{Nakane2003-1,Henrickson2000-1},
as well as what molecule is under investigation, and will
have repercussions on the ability to detect and sequence.
The translocation through the pore will be driven by the applied bias, but depends on
many factors, including the polymer-pore interactions, ionic effects, and viscous drag.
However, two properties are common to translocation in nanopores, namely the effective charge and screening of polymers within the pore, and the related issue of where the applied voltage drops.

In the absence of a polymer within a pore, one expects that the
majority of the voltage drop between the two ion chambers occurs in
the pore since it has a resistance higher than the surrounding
solution. However, depending on its shape (which depends
on whether the latter is biological or synthetic), one would expect
that the presence of a polymer in the pore can change significantly
the voltage drop of the system. For instance, with the
$\alpha$-hemolysin pore in Figure~\ref{alphapore}, one might expect
that when the polymer creates a much higher resistance, this occurs
primarily in the region of the pore which is roughly 2 nm in
diameter. Thus, the potential profile will change and drop more
significantly over the stem. This would cause the applied bias to
act most significantly on the nucleotides (or charged polymer units)
that are located within the stem.

There is some indirect evidence that this is the case.
\onlinecite{Meller2001-1} showed that there is a transition in velocity
dependence of the translocation process. Above
about 12 nucleotides, the velocity of the strands is independent of
their length, suggesting that the bias is pulling only on a finite
number of charges and is acting solely against the drag of the
polymer within the stem. Below about 12 nucleotides, the velocity
increases (in a highly non-linear way) with decreasing length. For
12 nucleotides, the contour length is approximately 5 nm, thus in
agreement with the pore stem length. This transition is explained by
a model of polymer translocation through a pore of finite
length~\cite{Slonkina2003-1}, which gives a transition in velocity
when the polymer's length is equal to the pore length.

\begin{figure}
\begin{center}
\includegraphics*[width=5.5cm]{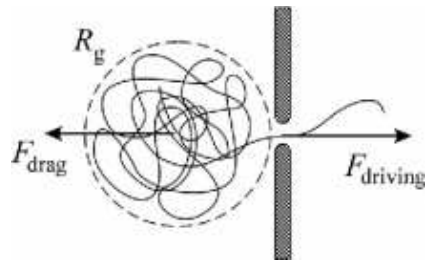}
\caption{Drag of a coiled-up polymer outside a pore.
From~\onlinecite{Storm2005-2}.}
%copyright American Chemical Society.
\label{Drag}
\end{center}
\end{figure}

Another interesting fact is that there is a transition in polymer
dynamics based on the pore size and pore-polymer interaction. For
pores with diameter roughly the size of the polymer width, one
expects strong interactions of the polymer with the walls of the
pore, thus strongly increasing its resistance to flow. When this is
the case, the timescale of polymer translocation, $t_d$, would be
controlled by this resistance, which would dominate over other
factors, such as polymer unraveling and drag outside the pore (see
Figure~\ref{Drag}). In this regime, one expects the translocation
duration to be linearly proportional to polymer length (for polymer
lengths much larger than the pore length). Indeed this is what has
been measured~\cite{Kasianowicz1996-1}. On the other hand, if the
pore is very wide (or there is very little polymer-pore
interaction), the translocation time can be controlled by other
factors, and one does not expect this time to have a linear relation
with the polymer length.

Such a nonlinearity has been observed and explained 
by~\onlinecite{Storm2005-2} and~\onlinecite{Storm2005-1}. These
authors observe that the (most probable) translocation duration
scales as
\begin{equation}
t_d \propto L^{1.27}
\label{DekkerScale}
\end{equation}
for unfolded ds-DNA translocation. By considering a model that
accounts for the driving electric force and the hydrodynamic drag
force on the coiled-up polymer outside the pore, \onlinecite{Storm2005-1}
obtain a translocation duration that scales with polymer length as
$L^{1.22}$, which compares well with the experimental value. The
unusual exponents come from the fact that the radius (of gyration)
of a polymer coil scales as function of its length with non-integer
exponent, and depends on the polymer type (self-avoiding, etc.). The
drag force will be dependent on the surface area of the coiled
polymer. Most sequencing technologies will probably have to deal
with small pore diameters, and thus in the regime where drag of the
polynucleotides within the pore is significant. Therefore, in these
devices the translocation duration should scale with the length of
the polynucleotide. More generally though, this result shows that if
one wants to use a nanopore as a polymer detector, the latter has to
be first calibrated to take into account specific details which
affect the measured characteristic signals (similar calibration has
to be performed in the case of sequencing~\cite{Lagerqvist2006-1},
see subsection~\ref{sec:seqcurr}).

The linearly extended polymer may not be the only one able to
translocate. If the pore diameter $D_p$ is large enough to
accommodate multiple strands, there could be ionic blockade events
corresponding to folded polymers. In this case, let us analyze the
forces on the DNA nucleotide pairs, and, for folded DNA, on the
nucleotides that fall within the pore. This analysis 
highlights the complexity of polymer dynamics in nanopores. If
the applied voltage drops solely over the pore, the driving force on
the polymer in a given region of the pore will be
\begin{equation}
F_{driving}(m)\approx mF_{driving}^o
\end{equation}
where $m$ is the number of folds in the region, and $F_{driving}^o$
is the driving force on a single nucleotide pair,
\begin{equation}
F_{driving}^o=z_{eff}E
\end{equation}
where $z_{eff}\approx -0.5 e$ (see Sec.~\ref{sec:seqforce}) is the
effective charge on the nucleotide pair and $E$ is the electric field. If
pore-polymer interactions are responsible for the majority of the drag
force, then one also obtains a similar relation
\begin{equation}
F_{drag}(m)\approx m F_{drag}^o \label{dragrel}
\end{equation}
where, e.g., $F_{drag}^o \approx \eta v$, with $\eta$ a coefficient that
is proportional to the surface area for contact of the polymer and
pore surface (that is the reason why $\eta \to m \eta$ for the
folded polymer), and $v$ is the polymer velocity. For a constant
polymer velocity\footnote{This, of course, assumes that there is
such a thing as a ``constant velocity'' for this nanoscale object.
This is not strictly true, and estimates of drag and subsequent
analysis may be affected by a varying velocity.},
\begin{equation}
F_{drag}=F_{driving}
\end{equation}
and, thus, the factors of $m$ will cancel, and $v$ would be the same
regardless of whether the polymer is folded or unfolded ($v=v_o$).
Now let us consider a polymer that has a region of length $L_m$ of $m$ 
folds, with $L_m \gg L_p$.
If we suppose volume exclusion to be responsible for ionic blockade
(see next section), the blockade current of the folded and unfolded 
polymer will be related by 
\begin{equation}
I_o-I_b^m=m(I_o-I_b^o)
\label{blockrel}
\end{equation}
where $I_o$ is the open pore current, $I_b^o$ is the blockade
current for an unfolded polymer, and $I_b^m$ is the corresponding
quantity for the polymer with $m$ folds.
Given Eq.~\ref{blockrel}, one finds that the total charge blocked for the translocation of 
the folded polymer region is equal to the
total charge blocked by a linearly extended region of length $mL_m$,
which contains the same number of nucleotides,
\begin{equation}
\left( I_o-I_b^m \right) \frac{L_m}{v} = \left( I_o-I_b^o
\right) \frac{mL_m}{v_o} \, .
\end{equation}
This is what has been dubbed ``the rule of constant event charge deficit''~\cite{Fologea2005-1}
\begin{equation}
\mathrm{ecd} \equiv \Delta I_b t_d=\mathrm{constant},
\end{equation}
where $\Delta I_b = I_o - I_b$, and $t_d$ is the translocation time.
In experiments where this holds, it would seem to justify an
excluded volume model for the blockade current.

This phenomenon has indeed been observed in experiments with 10 nm
diameter Si$_3$N$_4$ pore (5-10 nm thick) and 10 kbp
ds-DNA~\cite{Li2003-1}. A smaller Si$_3$N$_4$ pore of 4 nm diameter
also showed a set of events that were distributed according to this
``rule''~\cite{Fologea2005-1}.
%\footnote{Chen {\it et al}. found
%that the percentage of unfolded events increases with increasing
%voltage~\cite{Chen2004-1}, which makes sense due to entropic
%factors.}
However, these observations were only for larger pores (4
and 10 nm for ds-DNA, 4 nm for ss-DNA), and thus they are not
necessarily indicative of the physics of smaller pores.
There are also events
that do not fall on curves of constant ecd~\cite{Fologea2005-1}.
These seem to be events where the ds-DNA temporarily bond to the
walls of the pore or have some other factor controlling their
velocity (besides viscous forces), such as untangling of the ds-DNA
outside of the pore.

To get the ``constant event charge deficit,'' one has to assume that
a pore-polymer interaction or, at least, a polymer interaction
specific to the pore region (e.g., an electrostatic drag induced by
the confinement) is causing the drag which will give the relation in
Eq.~\ref{dragrel}. With this type of drag, one expects the translocation duration to scale as $L$,
the length of the polymer. However,~\onlinecite{Storm2005-2}
find that the (most probable)
translocation duration scales as in Eq.~\ref{DekkerScale}. This
seems to indicate that it is not the polymer-pore interaction creating
the drag but rather the externally jumbled polymer.
There are many possible physical explanations for such a discrepancy
and other experiments are needed to have a complete understanding of
these effects.

\subsubsection{Specifics}

\begin{figure}
\begin{center}
\includegraphics*[width=7.5cm]{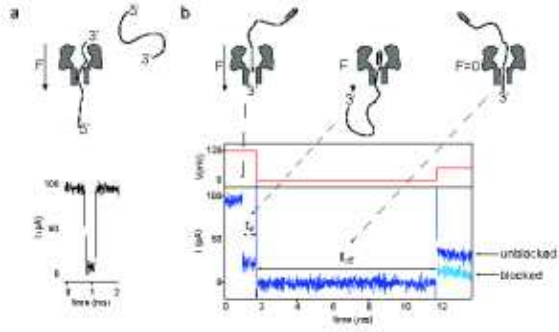}
\caption{Schematic of a hairpin experiment with an $\alpha$-hemolysin pore. (a) A typical
experiment on ss-DNA can pull the strand through from $3^\prime \to 5^\prime$
or from $5^\prime \to 3^\prime$ leading to blockades with different characteristics. (b) The
double-stranded portion of the DNA hairpin can not transverse the pore. Thus, the hairpin
can be pulled as far as possible into the pore, and one can turn off the pulling voltage for some
time $t_{off}$ after which one can check whether the strand is still in the pore by applying
some small probing voltage.
From~\onlinecite{Mathe2005-1}.}
%copyright US National Academy of Sciences.
\label{Hairpin}
\end{center}
\end{figure}

Here we give just one example of an interesting {\em specific} property
observed for PN experiments with $\alpha$-hemolysin pores. We
previously discussed results of~\onlinecite{Kasianowicz1996-1} (see
also~\onlinecite{Akeson1999-1}) which raise an intriguing question
about the directionality of translocation and capture of the PN. Are
these two processes dependent on which end, the $5^\prime$ or
$3^\prime$, translocates first? Recently, both experiments and MD
simulations have confirmed directionality-dependent translocation. 
\onlinecite{Mathe2005-1} performed experiments on DNA hairpins, as shown
in Figure~\ref{Hairpin} (see also~\onlinecite{Henrickson2000-1}). 
The experiment has been
performed by pulling the hairpin into the pore, measuring the ionic
blockade, then turning off the pulling voltage for a time $t_{off}$
and using a small probing voltage to detect whether the pore was
still blocked. These authors have found that the hairpins with the
$3^\prime$ end dangling (translocating $3^\prime \to 5^\prime$) has
a lower blockade current than the $5^\prime$ end (translocating
$5^\prime \to 3^\prime$), see Figure~\ref{Hairpin}. Further, these
two blockade currents are in good agreement with the two peaks found
from performing the typical single-stranded experiment (with no
hairpin). This indicates that early experiments obtaining
translocation events with two different blockade currents were
measuring the difference in orientation of the strands. In addition,
the diffusion and velocity (when pulled) of the hairpin out of the
pore was found to be much slower for the $3^\prime$ end entering,
e.g., when the strand is translocating $5^\prime \to 3^\prime$. This
also is a good indicator
 that the early observation of two lifetimes (with different capture rates)
of translocation events was likely due to different orientation of
the translocating DNA.

In addition, \onlinecite{Mathe2005-1} performed MD simulations which
demonstrated that in a pore the bases on the DNA tilt toward the
$5^\prime$ end. This explains the slower motion
of the DNA, because when moving $5^\prime$ end first there is
additional ``mechanical'' friction. This phenomenon will be
important not only for detection of different types of
polynucleotides, but also for any of the sequencing proposals
described below. For instance, if there is a different average base
orientation depending on which end enters first, the current
distributions of the bases (see subsection~\ref{sec:seqcurr}) may
also be different depending on whether transverse control erases the
directional differences. One may also ask whether this difference in
$5^\prime$ and $3^\prime$ orientations will be present in a
synthetic pore, or whether it depends on properties specific to the
$\alpha$-hemolysin pore. Further, the interplay between this effect
and that of secondary structure is still unclear, pointing to the
need for more experimental and theoretical investigation.
%\footnote{Meller {\it et al}. found that poly(dA) has two lifetime
%peaks, with the longer lifetime strongly dependent on temperature,
%coinciding with the strong temperature dependence of the secondary
%structure of poly(dA)~\cite{Meller2000-1}. Also, the fraction of
%events falling into the longer lifetime peak increases with
%decreasing temperature, as one would expect for a secondary
%structure. Further, the dispersion of the lifetime of this peak
%grows with decreasing temperature, corroborating the hypothesis of
%the break up of the secondary structure as it enters the pore.
%Finally, other polymers such as poly(dAdC), which do not have strong
%stacking (and therefore no significant secondary structure), did not
%show these same properties.}

\subsection{Pore and ion electronics}
\label{sec:ion}

From the above discussion we conclude that ionic currents through the
 nanopore are important for the detection of polynucleotides.
However, much is still not understood about the translocation of
ions themselves. The interplay between volume exclusion, steric
blockage, electrostatics, solution-ion dynamics, and interaction
with the pore itself is quite complex and poorly understood.

\subsubsection{Ionic and blockade currents}

In part, the blockade current forms due to ``volume exclusion''. The
polynucleotide occupies part of the pore's volume which partially
blocks ions from occupying the pore. This reduces the number of
charge carriers and thus the current. How much of the volume is
occupied by the PN depends on the structure and composition of the
strand. PNs with helical secondary structure could block more of the
pores area if the secondary structure stays intact. There are
additional questions on how much other factors, like, e.g., DNA-pore
interactions and electrostatics contribute to the blockade.

To get an idea of the contribution of volume exclusion to the blockade
current, we can use a steady state form of the Nernst-Planck (NP) 
equation~\cite{Chen1992-1,Barcilon1992-1,Coalson2005-1,Schuss2001-1}
\begin{equation}
J_z=e \mu n E_z - e D \frac{\partial n}{\partial z}
\label{NPeq}
\end{equation}
for the charge flux $J_z$ in the $z$-direction, where $e$ is the
electric charge (assuming monovalent cations), $\mu$ is the ion
mobility, $n$ is the charge carrier density, $D$ is the diffusion
coefficient, and $E_z$ is the driving field in the $z$-direction.
The general assumption that goes into the NP equation is that of a
continuous ionic distribution contained within a continuum dielectric.
The particular form of the Eq.~\ref{NPeq} assumes that there is no
free energy barrier to ion transport through the pore and thus it is a great start
to examine whether volumetric effects are responsible for the blockade current.

%If there is no free energy barrier to ion transport into the pore
%(or in a limiting sense, if the applied bias is very large, allowing
%us to ignore any free energy barrier), then there would not be ion
%buildup at the pore entrance and we can neglect the diffusion term.
%We are then left with only the first term in the equation which
%describes driven ions.
In the absence of any substantial diffusive term, the open pore current reduces to
\begin{equation}
I_o=e \mu n E_z
\end{equation}
where $E_z \approx V/L_p$ (the voltage $V$ drops mostly over the
pore length $L_p$). The quantity $e \mu n=\sigma$ is the
conductivity of the pore.

If volume exclusion were the sole factor in the change in current,
one could write,
\begin{equation}
n_b=\frac{V_p-V_N}{V_p} n = F n
\end{equation}
where $n_b$ stands for the carrier density during a blockade event,
$V_p$ is the pore volume for one repeat unit of the PN, and $V_N$ is
the volume of a nucleotide (see Table~\ref{sizes}). For the fraction
reduction in current from open ($I_o$) to blocked ($I_b$) pore we
then get
\begin{equation}
\frac{I_o-I_b}{I_o}=1-F \, .
\label{bleq}
\end{equation}
More simply, if one assumes that the carrier density is the same,
but that part of the pore area $A_o$ has been blocked, then
$(I_o-I_b)/I_o=A_b/A_o$, with $A_b$ the portion of the area which is
blocked. If we assume the same length $l$ for both the unblocked and
blocked pore, this relation will also give equation~\ref{bleq}.
The whole analysis requires a regularity in the
orientation of the blocking species, which may not be
completely correct.

For random coil (extended) ss-DNA, $F$ can be matched with the
volume fraction from Table~\ref{sizes}. The volume of the
nucleotides in Table~\ref{sizes} are for DNA nucleotides, but should
be essentially the same for RNA nucleotides, and the repeat units in
the PN, whether charged or passivated. This is because {\it i)} the
extra hydroxyl group has marginal effect on molecular volume, and
{\it ii)} the secondary structure should also not change the volume
occupied per nucleotide. The reason for the latter is simple: the
molecular volumes were computed with the van der Waals radii, which
are of the same order as half the base separation in stacked
secondary structures. For example, the van der Waals radius is 1.7
$\mbox{\AA}$, but the base separation is $\sim 3$ \AA. Thus, one
may expect that secondary structure can slightly increase or decrease the
nucleotide volume, but only by a small fraction.

Consider now random coil poly(C), or poly(dC), within the
$\alpha$-hemolysin pore. Let us take 15 $\mbox{\AA}$ as an estimate
for the pore radius, and suppose the PN is maximally linearly
extended such that each nucleotide occupies a 7 $\mbox{\AA}$ length
of pore.\footnote{This would give about 7 nucleotides in the pore stem, but 
there may be more depending on secondary structure.} In this case,
$1-F=0.26$ for cytosine, which does not come close to the blockade
values ($>90\%$~\cite{Akeson1999-1}). When the secondary structure
changes, this will change the length of the pore that each
nucleotide occupies. If, for instance, the bases stack at a distance
3.4 $\mbox{\AA}$ (poly(C) stacks at 3.1 $\mbox{\AA}$~\cite{Arnott1976-1}),
giving a minimum occupation length in the pore,
then $1-F=0.52$. This is still much less than the blockade values
found experimentally.

Of course, one may need to compute molecular volumes including the
hydration of the nucleotides~\cite{Deamer2002-1} and also divide the
pore into ``good'' and ``bad'' volume, as there may be bound water
molecules on the walls. This would certainly give a larger fraction
reduction in volume. However, defining volumes with the hydration layers included may not be accurate in the dynamical pore environment, as translocating ions may temporarily share hydration layers with the nucleotides. 
But this raises some
interesting questions about the interaction of the different bases
with the nearby water and ions. For instance, do the bases interact
sufficiently different with the water that the free volume,
including hydration layers, distinguishes the bases better than in
Table~\ref{sizes}, or is it the reverse? One would expect the
latter, because the hydration layers probably smooth over
differences in the bases, and the backbone is likely the main
location for bound water molecules.

Regardless, the argument seems strong enough that it points to other
effects (interactions with the pore walls and electrostatics) to be
the main cause of the blockade current for DNA. Other polymers, especially
neutral ones such as polyethylene glycol, may show different behavior.
However, the the blockade is still a complex phenomenon. For example, poly(C) and
poly(dC) have the same base, but the interaction with the pore walls and
dynamics could change due to the secondary structure. In this case,
the more structured poly(C) may create dynamical charge traps
within the pore (a steric effect), thereby reducing the current
further than for poly(dC). Also, poly(dC) creates larger current
blockages than poly(A) (89$\%$ and 86$\%$ for poly(dC) compared to
85$\%$ and 55$\%$ for poly(A)~\cite{Akeson1999-1}). Since C has a
much larger association with the lysine in the $\alpha$-hemolysin
pore~\cite{Bruskov1978-1}, interactions may be the determining factor.

Further, there are other effects to consider. Volume exclusion
creates an additional electrostatic barrier due to the increased
confinement of the ions. Thus a small change in volume can decrease
the current because the associated free energy change
``deactivates'' many ions for transport (see below). Additionally,
PNs remain charged within the pore~\cite{Keyser2006-1,Cui2007-1}. How does
this affect the blockage of coions, and the flow of
counterions?\footnote{An unusual phenomenon of current enhancement
due to DNA translocation was found in some
experiments~\cite{Chang2004-1,Fan2005-1,Chang2006-1}. This has been
explained in terms of an enhanced counterion current due to the
presence of the backbone charge of the DNA, which at low ionic
concentrations overpowers the volume exclusion.} We are not aware of
a complete answer to this question.

\subsubsection{Electrostatics}

Another issue related to the ionic currents is how the
electrostatic environment of the pore affects transport? Some partial answers 
can be found by looking at simplified models of the pore
environment, and using this as input to the full NP
equation.\footnote{One can also question the validity of that equation because
inhomogeneities at the nanoscale may be important. We are not aware of a critical study 
of this point.}
For instance, one could take an appropriate free energy potential, and
possibly position-dependent dielectric constants and mobilities. In the case
where the pore diameter is small, the free energy change can be
quite large and thus significantly suppress ion transport.

In the regime of large free energy change,
we consider a scenario where the two chambers have an equilibrium concentration
of ions at their respective shifted chemical potentials. Most of the ions do not have the energy
to transport across the pore. So we consider an activated process where the number of
ions available for transport is given by
\begin{equation}
n=n_o e^{-\Delta F_b/kT}
\label{act}
\end{equation}
where $n_o$ is the bulk density and
$\Delta F_b$ represents the free energy barrier due to both an electrostatic energy barrier
and an entropic barrier. For large pores, both of these contributions will be negligible as the
electric field from the ion is still embedded in a large volume of water and there is still
a large phase space for ions to occupy.

\begin{figure}
\begin{center}
\includegraphics*[width=5.5cm]{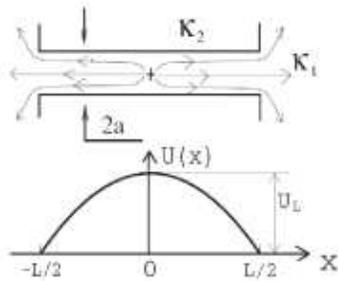}
\caption{Continuum pore model used by~\onlinecite{Zhang2005-1}. The dielectric
constants are such that $\kappa_1 \gg \kappa_2$ so that all field lines of a charge in the
pore run down the pore axis. The lower panel shows the electrostatic energy barrier for a
single charge to transport from the bulk into the pore.
From~\onlinecite{Zhang2005-1}.}
%copyright American Physical Society
\label{Model}
\end{center}
\end{figure}

For smaller pores, one can think of the simplified model shown in
Figure~\ref{Model}. The general concept is to look at the pore as a
continuum large dielectric (water) region of space surrounded by a
material of low dielectric (SiO$_2$,Si$_3$N$_4$), thus creating an
electrostatically quasi-one-dimensional
structure~\cite{Zhang2005-1,Teber2005-1}. Under this simplified
picture, the electrostatic and entropic components to $\Delta F_b$
can be computed. \onlinecite{Zhang2005-1} found that this simplified model
gives a tremendously suppressed ionic current due to the large free
energy change of ions entering the pore. However, they also found
that any significant presence of wall charges, which are present in
experimental systems such as silicon dioxide, would
decrease the free energy barrier and thus increase the ionic
conductance~\cite{Zhang2005-1}. Many recent papers have looked at
nanopore-ion-DNA electrostatics from this point of
view~\cite{Zhang2005-1,Zhang2006-3,Kamenev2006-1,Zhang2006-2,Bonthuis2006-1}.

\begin{figure}
\begin{center}
\includegraphics*[width=7.5cm]{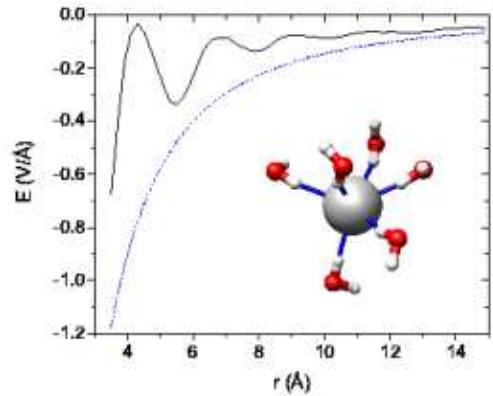}
\caption{(Color in online edition) Microscopic electric field of a
single anion in a spherical water droplet of radius 25 $\mbox{\AA}$
(solid black line). The dashed blue line is the bare field. The inset
shows a model of the first hydration layer.} \label{field}
\end{center}
\end{figure}

In our own work~\cite{Lagerqvist2007-1}, we have looked at the
microscopic effects of the electric fields and solution structure.
If one examines the field of a single anion surrounded by a
spherical droplet of water, this looks very different than one would
expect from a bulk dielectric.\footnote{This is, of course, due in
part to the fact that macroscopic electrostatics requires averaging
the electric field over large regions of space of at least 10
nm~\cite{JacksonWater}. It is also due to the formation of the
hydration layers as a consequence of the strong local electric
field.} The field obtained from MD simulations, and averaged over
time and solid angle, is shown in Figure~\ref{field}. One can
clearly see the formation of a nanoscale structure around the ion: the
water molecules start to form layers (called {\em hydration
layers}). For instance, for the Cl$^-$ we show here, there are on
average six water molecules in the first layer which orient their
dipoles very strongly toward the anion. These hydration layers only
get partially destroyed within a pore~\cite{Lagerqvist2007-1}, and
the free energy barrier has an electrostatic contribution given by
the partial destruction of the hydration layers. As the pore radius
is increased, there is a critical radius that allows a full
hydration layer to be transported through the channel. This creates
a strong non-linear dependence of the free energy barrier as a
function of pore radius and also ionic concentration. Effects like
the above illustrate the need to take a microscopic approach to
understanding nanopore electronics, and foreshadows novel physical
phenomena that will be observed at the interface between solids and
liquids (and biomolecules) in the nanometer regime.

\section{Sequencing and detection}
\label{sec:seq}

\setcounter{footnote}{0}

We have reviewed the physical characteristics of the bases and
polynucleotides, and also discussed nanopores and DNA translocation.
Thus, we are now in a position to examine in more detail the physical mechanisms by
which DNA can be detected and sequenced. Very loosely, we divide the
physical mechanisms into three categories: electronic, optics, and
force methods. In
this section we will discuss several of the proposed methods in
terms of the physical differences of the bases, and outline any
experimental and theoretical results pertinent to either measuring
these property differences or constructing a working apparatus.

%Most proposals for DNA sequencing have yet to be experimentally
%demonstrated. Simulations have thus been critical to examine the
%feasibility of proposed methods, and understand the fundamentals of
%device operation. Further down the road, simulations will be
%important to suggest new experimental configurations and materials,
%as well as possible modifications and improvements (like the
%integration of biological sensors and synthetic pores/materials).

\subsection{Electronic}
\label{sec:seqelect}

\begin{figure}
\begin{center}
\includegraphics*[width=7.5cm]{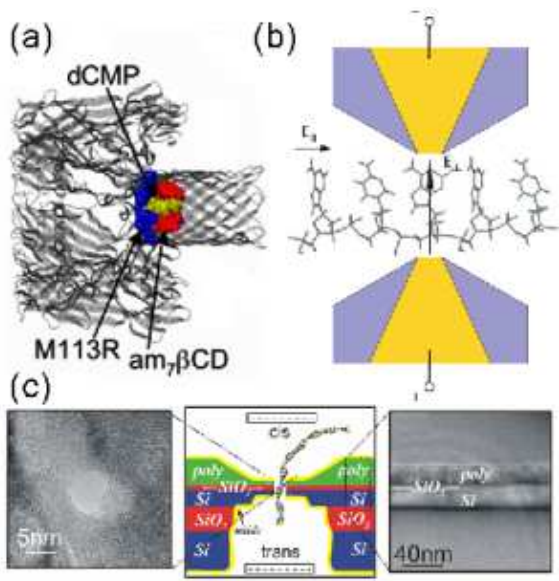}
\caption{(Color in online edition) (a) Engineered $\alpha$-hemolysin
pore used to detect nucleotides via their ionic blockade.
Adapted from~\onlinecite{Astier2006-1}. (b) Schematic of two electrodes
embedded in a nanopore. As PN translocates through the pore, the
electrodes drive a current across the nucleotides in the transverse
direction. The current across each of the different nucleotides
provides an electronic signature of the base. E$_\parallel$ is due
to the voltage pulling the DNA through the pore. E$_\perp$ is the
field perpendicular to the electrode surface and is provided by the
applied voltage across the electrodes and any additional external
capacitor. From~\onlinecite{Lagerqvist2006-2}.
%copyright Biophysical Society??
(c) Middle panel: schematic of a nanopore through a capacitor made
of doped polycrystalline silicon-SiO$_2$-doped crystalline silicon. The SiO$_2$
can be as thin as 0.7 nm~\cite{Gracheva2006-1}, which is about the spacing of nucleotides
in extended ss-DNA.
From~\onlinecite{Gracheva2006-1}.}
%copyright IOP
\label{combo}
\end{center}
\end{figure}

The electronic methods proposed so far are based on the ionic
blockade current in the nanopore~\cite{Kasianowicz1996-1,Deamer2002-1}, embedding nanoscale
electrodes in the nanopore to measure transverse transport across ss-DNA~\cite{Zwolak2005-1,Lagerqvist2006-1,Lee2007-1},
or measuring the voltage fluctuations in
a capacitor across the longitudinal direction of the pore~\cite{Gracheva2006-1,Heng2005-1}.
Three specific proposals using these techniques are shown in Fig.~\ref{combo}.

\subsubsection{Ionic blockade}
\label{sec:seqion}

With the promising first results of~\onlinecite{Kasianowicz1996-1},
it was speculated that it may be possible to sequence DNA by
measuring the ionic blockade in a very particular pore under the right conditions
(see~\onlinecite{Kasianowicz1996-1},~\onlinecite{Deamer2002-1},
and~\onlinecite{Deamer2000-1}). With this idea in mind, researchers
started to examine what kind of information can be extracted
from the nanopore experiments. Indeed, other experiments with
$\alpha-$hemolysin pores have shown that the ionic blockade can be
used to detect blocks of nucleotides, and that the blockade characteristics give
information on secondary structure and directionality ($5^\prime$,$3^\prime$)
of PN translocation.

For instance, by looking at RNA homopolymers poly(A), poly(U),
and poly(C), ~\onlinecite{Akeson1999-1}
have demonstrated that the blockade current and the translocation duration
can be used to distinguish between the type of base present in the homopolymers.
At 120 mV bias, it was found that
poly(U) translocation takes 1.4 $\mu$s/nt and 6 $\mu$s/nt for
two translocation event types compared
to 22 $\mu$s/nt for poly(A). Yet, they both can give blockade currents of essentially
identical magnitude (approximately 85$\%$ blocked). On the other hand,
poly(C) was found to give 95$\%$ and 91$\%$ blockade of the ionic current, and to
translocate at $\sim 5$ $\mu$s/nt.

An interesting feature of the above results is that poly(C) gives a
larger blockade of the ionic current than poly(A) even though it is
a smaller base, see Table~\ref{sizes}. \onlinecite{Akeson1999-1} have
suggested that this could be due to the secondary structure of
poly(C), which at neutral pH and room temperature has a helical
structure that is 1.3 nm in diameter~\cite{Saenger1988-1}. This size
is small enough to fit into the $\alpha$-hemolysin pore without
unraveling. To test this, the above authors have compared poly(C) to
poly(dC), its DNA counterpart. Poly(dC)
secondary structure is not as stable as poly(C), so differing
results for poly(dC) and poly(C) would give evidence for secondary
structure as the cause. \onlinecite{Akeson1999-1} indeed found that
poly(C) blocked more ionic current than poly(dC). Thus the
secondary structure of poly(C) is the likely cause of the larger
blockade. This, however, leaves an open question raised earlier in section~\ref{sec:ion}:
why does the secondary structure cause a larger blockage of current?

The slow translocation of poly(A) is also thought to be due to
secondary structure. Poly(A) has a secondary helical structure,
that, unlike poly(C), is too big to translocate through the
pore~\cite{Akeson1999-1}. Thus, poly(A) takes extra time to unravel
and go through the pore. On the contrary, poly(U) does not have any
secondary structure under the reported experimental conditions (see
discussion in~\onlinecite{Akeson1999-1} and~\onlinecite{Kasianowicz1996-1}).

It has also been shown that a PN with two homogeneous blocks,
A$_{30}$C$_{70}$, gives ionic blockade events with a stair
structure: a higher current is present when the homogeneous block of
A's is present in the pore, but this current decreases when the C's
are in the pore~\cite{Akeson1999-1}. Although far from
single-nucleotide resolution, this shows that one can obtain some
internal information about the strand from the nanopore. This result
together with using the average blockade current, average
translocation time, and the temporal dispersion~\cite{Meller2000-1}
to distinguish individual molecules, shows some of the promise of
nanopore technology in polynucleotide detection.

Later, \onlinecite{Fologea2005-1} took advantage of the ability of
solid-state pores to work under many conditions by examining DNA
translocation at many different pH's in order to detect the ds-DNA
denaturation into ss-DNA. As they increase the pH from 7 to 13,
these authors find that there is a rather abrupt transition when the
current blockade drops by roughly a factor of 2~\cite{Fologea2005-1}. That this is due to
denaturation is confirmed by measurement of the optical absorbance
which shows that denaturation occurs at around pH=11.6. These are
examples of physical processes that can be studied with, and
detection capabilities that can be achieved by, nanopore
technology.

In order to actually sequence a strand of DNA single-base resolution
is required. In this case, the different sizes of the bases, and
the different interactions between the bases and pore, have to be
detected.
 For the direct use of the
ionic current as an indicator of the base present at a location in a
polynucleotide (i.e., for sequencing), the nanopore must have
a length $L_p$ of roughly one nucleotide ($< 1$ nm) and diameter somewhere in between
1 nm and 2 nm.
%\footnote{Another possibility for direct use of the
%ionic blockade current as a way to distinguish between bases would
%be to first perform a chemical amplification of the base present at
%each location, i.e., every single nucleotide in a given sequence is
%``amplified'' to $n$ nucleotides in a row, as is done in the optical
%sequencing technique in Sec.~\ref{sec:seqoptic}. This procedure is
%not necessarily guaranteed to work. However, in light of the
%experimental results in distinguishing homogeneous blocks of bases,
%one would expect it to work
%even in the presence of secondary structure. Although, it would
%potentially add time-consuming and costly steps into the
%sequencing procedure.}

If the ionic blockade is solely due to excluded volume, then one
would expect that the differences of the bases in a 1.5 nm diameter,
0.7 nm long pore would be just a couple of percent, as shown in
Table~\ref{sizes}. The noise on the current itself, both
intrinsically (due to ionic fluctuations) and due to structural
fluctuations of the nucleotides is likely to be much larger than
this.\footnote{ Molecular dynamics simulations
of~\onlinecite{Aksimentiev2004-1} indicate that fluctuations in the
ionic current due to structural changes are larger than the
differences between bases. These structural changes are in part
caused by interaction with the pore surface.} For instance, the
fluctuations of the ionic blockade current for homogeneous sequences
is about 30$\%$ of the average current (see Figs.
in~\onlinecite{Meller2000-1}). The noise for a single base should be
larger than for the sequences. Intuitively, comparing to an
$\alpha$-hemolysin pore with about 10 bases in the stem, one expects
the noise for one base to be larger by a factor of about
$\sqrt{10}$. This would mean noise on the order of a 100$\%$ of the
average current. There may be possibilities, such as changing
conditions like temperature, pH, etc., to minimize secondary
structure effects and make the circumstances more amenable to
sequencing. But even in the best case scenario, where only the
pore-located nucleotide is controlling the ionic blockade current,
it may not be possible to successfully distinguish the bases. See
further the arguments presented in Sec.~\ref{sec:ion}.

Nonetheless, more recent experiments have shown that single
base differences can be detected in strands. For instance,
\onlinecite{Vercoutere2001-1} have looked at several different length DNA
hairpins that contain no single-stranded
portion. The hairpins initially block the
vestibule part of the $\alpha$-hemolysin pore, and would only be
pulled through the pore when the double strand temporarily unravels.
These authors have found that one can discern two hairpins with only
a difference of one base in the loop and also two hairpins with only
a mismatch as a difference. Both the vestibule blockade current and
the vestibule blockade duration were found to be different.
The hairpin with the base pair
mismatch is much more likely to unravel, and therefore would spend
much less time in the vestibule before translocation (a factor of
100 less time according to these experiments). Similarly for the
shorter loop hairpin, which is under much more strain.

Using ionic blockade, the same group has looked further at hairpin
differences, computer-learning algorithms for distinguishing
signatures, and the physics of hairpins within the
pores~\cite{Vercoutere2001-1,Vercoutere2003-1,Winters-Hilt2003-1,DeGuzman2006-1}.
Another group obtained similar findings for a different type of
hairpin experiment, where a single base difference in a
single-strand leg of a hairpin could be
detected~\cite{Ashkenasy2005-1}. It is important to note, however,
that these experiments do not actually achieve single-base
resolution as required for sequencing.

Without amplification of the bases, it is unlikely that a
strand of bases can be sequenced with the
bare ionic current.
However, there is potential to go beyond the normal current blockade experiments
by creating designer pores~\cite{Siwy2005-1,Bayley2001-1}. For instance, introducing a foreign
molecule into the pore or by using a pore that has some
chemically-specific affinity, the pore can interact in a particular
way to each of the four bases of DNA. A specific interaction could,
for instance, create a different level of pore blockade for each of
the bases. In this way, a distinguishable ionic current may be
obtainable. The underlying premise was demonstrated some years ago
by \onlinecite{Gu1999-1} with the use of an ``adapter'' molecule in a
$\alpha-$hemolysin pore. The adapter molecule was
non-covalently inserted into a pore and it helped detect organic
molecules that go through the pore by specific binding interactions
which change the value of the current blockade and the translocation
time.

\begin{figure}
\begin{center}
\includegraphics*[width=7.5cm]{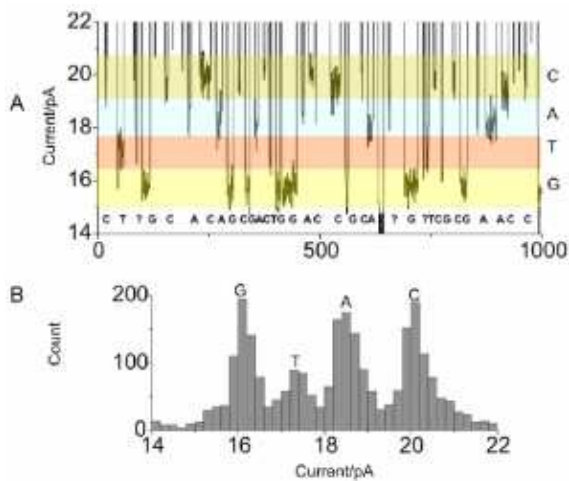}
\caption{(Color in online edition) (A) Current trace with all nucleotides present in solution (time is in milliseconds).
(B) Current distributions from the engineered pore.
From~\onlinecite{Astier2006-1}.}
%copyright American Chemical Society
\label{IonCurrDist}
\end{center}
\end{figure}

A very recent article has gone a long way in demonstrating
the use of a molecular adapter in
sequencing~\cite{Astier2006-1}.
%\footnote{Such a possibility has also
%been explored
%in~\onlinecite{Howorka2001-1},~\onlinecite{Siwy2005-1},
%and~\onlinecite{Kohli2004-1}}
These authors have used a mutant
$\alpha$-hemolysin pore, see Fig.~\ref{combo}(a), with a positively charged cyclodextrin
adapter. The adapter can bind and unbind from the pore (it is not
covalently bound), creating a stochastic signal of successive ``on''
and ``off'' current blockades. The addition of nucleotides changes
this stochastic signal, creating additional events with a smaller
blockade current when the nucleotide binds to the adapter. These
smaller blockade currents are significantly different for the
different bases, as shown in Fig.~\ref{IonCurrDist}. Although the
apparatus is chemical, it is really detecting a physical difference
between the nucleotides. The difference between the bases
results in ionic current distributions with partial overlap,
which gives the ability to correctly identify a base to about
93-98$~\%$ accuracy. Thus, multiple measurements of each base in a
sequence are necessary. Unlike the transverse transport ideas we
discuss below, multiple measurements may not be a controllable feature of
the procedure and therefore the present technique may require
re-sequencing each strand, or using some sort of oscillating voltage
to pull the nucleotide in and out of the adapter to obtain the
desired accuracy~\cite{Dimarzio2003-1,Kasianowicz2002-2,Astier2006-1}.

A full sequencing procedure might employ exonuclease digestion to
take a DNA strand and remove a base at a time~\cite{Astier2006-1}.
The pore itself would have to have the exonuclease attached so that
each base is released into the pore and its almost unique blockade
signature detected. In addition, in order to actually couple the
exonuclease digestion with the pore detection, one has to deal with
the fact that the adapter is not covalently attached to the pore and
can unbind. Also unknown is whether the nucleotides actually pass
through the pore or whether they just bind and unbind from the
adapter staying in the cis chamber (where they are added). Further,
the exonuclease could create the rate-limiting step. For example, a
fast digestion rate of 1000 nt/s has been
observed~\cite{Matsuura2001-1}, which is still much slower than the
intrinsic speed of the nanopore experiment, on the order of $10^6$
nt/s. However, the rate will also depend on the binding timescales and
the bandwidth of the ionic current measurement.

It is clear from the experiments described in
this and earlier sections (as well as results not discussed~\cite{Wang2004-1}), that the
detection capabilities and the ability of ionic current experiments
to probe physical processes at the nanoscale is a promising field.
The use of molecular adapters to create a working sequencing method
is also promising, but there may need to be additional advancements
to increase its intrinsic speed.

\subsubsection{Transverse electronic current}
\label{sec:seqcurr}

When measuring the ionic current the size, secondary structure, and
base-pore interactions distinguish the bases. One can also measure
intrinsic electronic properties of the bases by, for instance,
embedding electrodes within a nanopore to measure the transverse
current through ss-DNA as it translocates through the
pore~\cite{Zwolak2005-1}. This method essentially detects the
electronic structure of the bases (i.e., the degree of
delocalization, shape, and energies of the bases' electronic states)
relative to the specific electrodes. A schematic of the idea is
shown in Fig.~\ref{combo}(b). We stress that transverse transport is
different than the (also interesting and yet not fully solved)
problem of longitudinal transport in
DNA~\cite{Endres2004-1,Diventra2004-1,Porath2004-1}. The latter is
another complex problem involving a disordered quasi-one-dimensional
molecule. Further, measuring transverse currents in other scenarios
has been suggested as a way to explore base differences and new
physics~\cite{Macia2005-1,Apalkov2005-1}.

Initial calculations on individual nucleotides ideally configured
between two nanoscale gold electrodes\footnote{Note that the
electrodes need not be made of gold. For instance, Golovchenko and
co-workers are pursuing the same idea with nanotube electrodes. The
nanotube electrodes are likely to be less reactive with the liquid
and solid environment and may thus be more stable than metals
commonly used for contacts. One may also use the nanotube as part of
other detection apparati~\cite{Meng2007-1}.} (see
Fig.~\ref{DOS}(b)), indicate that the different bases can give quite
different currents under some conditions~\cite{Zwolak2005-1}.
However, the different currents are very sensitive to the electrode
spacing, and also vary greatly if the nucleotides have complete
freedom in orientation. For a 1.5 nm electrode spacing and the bases
standing upright with respect to the electrodes (Fig.~\ref{DOS}(b)),
the electronic currents for the nucleotides are different by orders
of magnitude, except for the nucleotides G and C which differ only
by about a factor of 2. The reason for this large difference is the
coupling of the bases to the electrodes~\cite{Zwolak2005-1}. Due to
the different sizes of the bases, as shown in
Section~\ref{sec:phys}, their molecular orbitals have a different
spatial extent. The largest base, adenine, thus couples better to
the upper electrode. More recently the relation between base
orientation/size and conductance was shown explicitly
in~\onlinecite{Zhang2006-1}. This is both an advantage, in that it
gives added differentiation between the bases, and also a disadvantage,
because any fluctuations in the orientation of the nucleotide can
drastically change the current.

To quantify this, let us look at the tunneling current through a
single energy level $E_N$ in the limit of zero bias $V$.
For weak coupling, as is the case here, this is
given approximately by
\begin{equation}
I\approx\frac{e^2}{\pi\hbar}\frac{\Gamma_L(E_F)\Gamma_R(E_F)}{(E_F-E_N)^2}
V.
\end{equation}
This equation shows that the current is proportional to the two
coupling strengths $\Gamma_L$, $\Gamma_R$, and inversely
proportional to the distance between the energy of the state
and the Fermi level $E_F$. The different bases have nearly the same
energy levels compared to the Fermi level (see Fig.~\ref{DOS}),
thus this portion is not likely to provide
much distinguishability. However, the coupling elements can be very
different for the bases because of their different spatial
extensions and wavefunctions (see Sec.~\ref{sec:elect}).

A preliminary look at the effects of changing orientation showed
that the transverse currents are still quite different except for G
and C, thus bolstering the case for sequencing~\cite{Zwolak2005-1}.
Also, unlike the voltage fluctuation measurements discussed below,
the transverse current is minimally affected by nearest-neighbor
bases because the current is controlled by base-electrode coupling
and the energy of the molecular states. These are only slightly
influenced by the neighboring bases, so long as the electrode width
is on the order of the base spacing ($\sim7\mbox{\AA}$ for
extended ss-DNA). The base-electrode coupling is particularly
important, and this drops exponentially with the distance of the
bases from the electrodes. This is unlike the capacitance method we
discuss below where the Coulomb interaction is long
range\footnote{At these atomic scales, and in particular inside the
pore, screening by the few water molecules present around the bases
is not as effective as one would obtain by assuming the macroscopic
dielectric constant of water~\cite{Lagerqvist2007-1}} and thus
nearest neighbors can influence the voltage signal. However, at such
an electrode spacing of 1.5 nm the current is already
very small, on the order of picoamps for the base A. It is thus
necessary to go to smaller electrode spacing and also examine the
full effect of structural fluctuations.

To obtain a more realistic look at the effect of structural
fluctuations the two authors together with J. Lagerqvist have
performed molecular dynamics (MD) simulations of ss-DNA being pulled
through the pore~\cite{Lagerqvist2006-1}. In the absence of any
control on the DNA, the orientation of the bases varies wildly (the
bases can be at any orientation, perpendicular or parallel to the
electrode surfaces). This causes orders of magnitude fluctuations in
the value of the current. Due to these large fluctuations and small
values of the current magnitude, it is thus unlikely that sequencing
will be possible in the absence of any control on the DNA
translocation.

However, MD simulations show that a relatively weak transverse field
can orient the nucleotide in the junction with respect to the
electrodes in less time (in the 100's of picoseconds) than it takes the
nucleotide to translocate through the
junction~\cite{Lagerqvist2006-1,Lagerqvist2006-2}. Referring to
Fig.~\ref{combo}(b), the condition on the transverse field to be
strong enough is E$_\parallel \ll$ E$_\perp$. This ensures a slow
enough DNA velocity to allow the transverse field to orient the
nucleotides while they are facing the electrodes. The transverse
field strength can be of the same order of magnitude as that driving
the electronic current, but may be provided by an external
capacitor around the nanopore device.

\begin{figure}
\begin{center}
\includegraphics*[width=7.5cm]{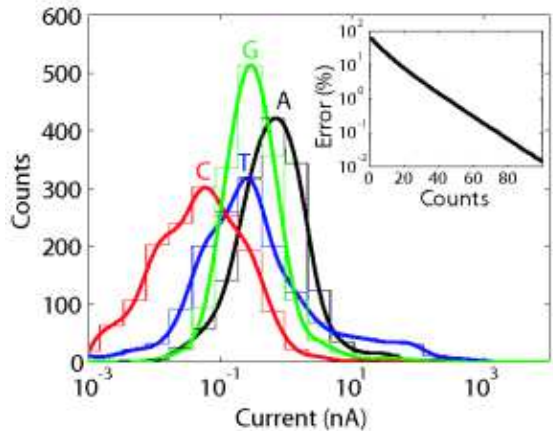}
\caption{Current distributions for the different nucleotides in a
1.25 nm electrode spaced pore and at 1 V, with a transverse
controlling field. The inset shows the error of misidentifying a
base versus the number of independent counts (or measurements).
From~\onlinecite{Lagerqvist2006-1}.}
%copyright American Chemical Society
\label{currdist}
\end{center}
\end{figure}

When the control is exerted on the nucleotides, one obtains current
distributions for the different bases that are sufficiently
disjoint, see Fig.~\ref{currdist}, so as to allow the bases to be
statistically distinguishable. In this case, the statistics can be gathered both intrinsically
and extrinsically. First, the strand has some finite velocity, thus
allowing for multiple measurements to be gathered as the nucleotide
translocates through the junction region (so long as the
 inverse of the bandwidth of the electrode probes is much smaller than
the translocation time of a nucleotide). Second, and more
importantly, the finite bandwidth of the probes itself samples over
many configurations of the intervening nucleotide.

Overall this leads to a sequencing protocol where one first has to
measure the current distributions for each of the nucleotides using
a homogeneous strand. These distributions will depend on the
particular pore geometry and conditions and thus will be particular
to each device. Then one can sequence the DNA by pulling it at a
velocity that allows each nucleotide to stay in the electrode region
for enough time to collect a current distribution. Then by comparing
with the ``target'' distributions, one can determine the
sequence~\cite{Lagerqvist2006-1}. To get an idea about the potential
speeds achievable with this method, we estimated that a single run
through the 3 billion bases on a single-strand of human DNA would
take a raw time of about 7 hours, without
parallelization~\cite{Lagerqvist2006-1}.

There are other issues yet to be explored, some of which have to be
addressed experimentally. For instance, the construction of a
nanopore with embedded electrodes remains a formidable challenge to
the implementation and testing of the method described above.
However, proof of principle of the distinguishability of the bases
via transverse current measurements can be done with other methods,
such as using a scanning tunneling microscope. This route has been
explored in the work of~\onlinecite{Xu2007-1,Xu2007-2}. These authors have
shown experimentally that the four bases (using isolated
nucleotides) provide a distinguishable signal via their HOMO level
when measured with an STM. Much earlier, other groups have also
examined isolated bases or bases in a strand of DNA with scanning
probe techniques (\onlinecite{Tao1993-1,
Tanaka2003-1,Hamai2000-1,Tanaka1999-1,Wang2001-1}). The idea of
using transverse transport to sequence is being pursued
experimentally by a few groups (M. Ramsey {\it et al}.~\cite{NHGRI},
Branton and Golovchenko~\cite{Golov}, J. Lee {\it et
al.}~\cite{Lee2007-1}).

There are, of course, drawbacks to the transverse transport approach
we have described. Clearly, if the amount of noise is large, it will
make the signals from the bases potentially indistinguishable. There
will be $1/f$ noise and noise provided by ionic fluctuations
(discussed below). While one can reduce $1/f$ noise by operating
away from zero frequency, the role of ionic noise is less clear.
Also, one may wonder in a real implementation whether clogging of
the pore can occur, i.e., whether the small size of the pore and the
transverse electric field(s) will cause too much interaction of the
DNA with the surface of the pore. This is an issue that has yet to
be investigated: its study requires a fully quantum mechanical
treatment of the DNA motion in the pore.

One very important and interesting issue, both fundamentally and for
the realization of the above approach, is the effect of ionic
fluctuations on electronic transport. It has
importance for many cases of charge transport in soft materials,
such as longitudinal transport in
DNA~\cite{Diventra2004-1,Endres2004-1,Porath2004-1}. It would also be
interesting to investigate what happens to the tunneling current in
the presence of fluctuating energy landscapes created by classical
objects (e.g., ions) that cross the electrons' path, when the rate of
these fluctuations is comparable to the inverse of the coherence
time. An interesting experiment in this direction may be to use
different weight counterions to observe the transition from slower
to faster fluctuations. The effect of ions on transport is part of the more 
general problem of understanding the effect of changing a single atom in molecular 
junctions~\cite{Diventra2000-1,Yang2003-1}.

Finally, another geometry has been proposed to probe electrical
transport perpendicular to the base {\em planes} (i.e., with the base
plane parallel to the electrode surfaces)~\cite{Lee2007-1}. However,
there are two challenging issues with this setup. For one, since the
transport is now envisioned across the base planes, the electrode
spacing has to be very small, $< 8 \mbox{\AA}$, to obtain a
measurable tunneling current through the bases. On the other hand,
this will make it difficult for the ss-DNA to translocate through
the pore. From our MD
simulations with selected initial conditions~\cite{Lagerqvist2006-1,Lagerqvist2006-2}, the smallest
pore that will allow DNA to translocate is $10 \mbox{\AA}$ in
diameter. In addition, even with an electrode spacing that allows
for a measurable current, the differences between the bases in the
above planar configuration are not as large in magnitude as in the
case in which the bases stand upright with respect to the electrode
surfaces (as shown in
Fig.~\ref{DOS}(b))~\cite{Zwolak2005-1,Zwolak2004-1}. This is due to
the fact that the bases have very similar HOMO and LUMO charge
distributions when viewed perpendicular to their plane, whereas when
standing upright, they have different couplings (see
Fig.~\ref{states}).

\subsubsection{Capacitance}
\label{sec:seqcap}

The measurement of voltage fluctuations in a metal-oxide-silicon
capacitor combined with a nanopore has been proposed as a method to
detect and obtain the length of DNA, and potentially to sequence
it~\cite{Heng2005-1,Gracheva2006-1}. A schematic of a nanopore
capacitor is shown in Fig.~\ref{combo}(c). As each nucleotide passes
through the pore, the charge on its backbone can induce a voltage
across a capacitor in the longitudinal direction. By measuring these
voltage fluctuations, one can effectively count the number of
nucleotides. For such a scheme to be used for sequencing it would
have to also be sensitive enough to detect the dipole moment
differences of the nucleotides given in Section~\ref{sec:phys}.

\onlinecite{Heng2005-1} have indeed observed voltage signals on the
two doped-silicon electrodes as a strand of DNA passes through the
pore, the difference of which gives oscillations. The oscillations
have a magnitude that could be adequate to detect the dipole
difference of Adenine and Thymine~\cite{Heng2005-1}. However,
experimentally these authors were not able to resolve even the
charge on the individual nucleotides because of the bandwidth and $RC$ time
constant of the probe, and the large size of the pore.

\begin{figure}
\begin{center}
\includegraphics*[width=7.5cm]{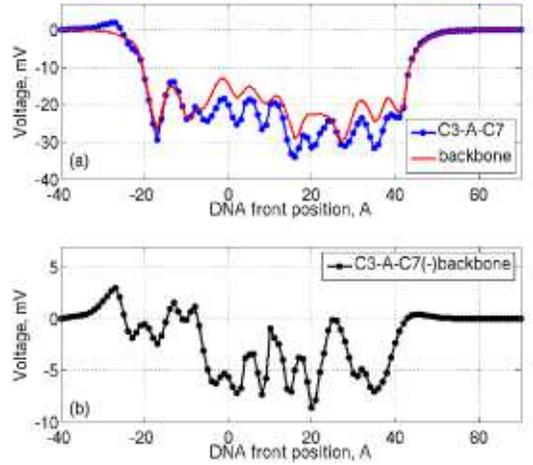}
\caption{(Color in online edition) An example voltage signal obtained from MD
simulations/electrostatic calculations of a 11 base single-strand of
DNA translocating through a nanopore capacitor. The dips correspond
to the nucleotides. Experimental signals have been obtained of the
voltage fluctuations, but without the fine resolution of these
simulated signals~\cite{Gracheva2006-1}.
From~\onlinecite{Gracheva2006-2}.}
%copyright IOP
\label{CapSig}
\end{center}
\end{figure}

The same group also performed 
simulations~\cite{Gracheva2006-1,Gracheva2006-2}, to calculate the
expected voltage fluctuations on the capacitor as a ss-DNA is pulled
through a 1 nm radius pore. In these simulations, the DNA strand was
not allowed to have conformational fluctuations (similar in style to
the calculation shown in Fig. 1 of~\onlinecite{Lagerqvist2006-1}).
An example signal obtained from this simulation is shown in
Fig.~\ref{CapSig}. Interestingly, the maximum signal obtained is
$\sim 35$ mV for the nucleotide, 30 mV for the backbone, and 8
mV for the base. In addition, as a nucleotide moves through the pore
the corresponding voltage signal has been found to be influenced by
up to three nearby nucleotides. Even though the bases themselves
show different voltage signals, the larger backbone signal seems to
dominate and give almost identical signals for the different
nucleotides (see Fig. 3 in~\onlinecite{Gracheva2006-2}).

Another important factor in the capacitance approach is the
dimension of the pore. Estimates suggest it should have about 1 nm
radius in order to obtain an adequate
signal~\cite{Gracheva2006-1,Gracheva2006-2}. The 1 nm radius of the
pore would both force the DNA to be stretched as it goes through the
pore, thus aligning the bases perpendicular the electrodes (which
produces a better signal than if the plane of the bases were
parallel to the electrode surfaces) and maximizing their distance
from each other (for instance, for the sequences
in~\onlinecite{Gracheva2006-2} the bases are about 7 $\mbox{\AA}$ apart,
double their distance than in ds-DNA or in ss-DNA with strong
secondary structure). Also, the small pore excludes a lot of water,
thus reducing screening.

Theoretically, though, one wonders how much, in principle, the bases
can be distinguished via their dipole moments. From
Section~\ref{sec:phys}, the dipole moments of C and G isolated bases
differ by just a percent (the differences between A and T, and
those with G,C, are much larger). This, however, ignores the effect
of the backbone dipole and the ion-counterion dipole, which are both
very large. In addition, the large fluctuations of ions at the
entrance of the pore may contribute substantially to the voltage
fluctuations. Indeed, most of the voltage drop occurs at the
location of the (high-resistance) pore. Thus, ions of opposite
charge would build up on the two open sides of the pore entrance.
When DNA enters the pore, it creates an even higher-resistance pore.
As a consequence, there should be a fluctuation in the ionic
concentrations near the pore ends, thus creating a temporary voltage
fluctuation on the capacitor. The consequence of this effect on
sequencing, together with conformational fluctuations of the DNA
bases inside the pore, has not been studied yet.

 To get some perspective on how
meaningful the dipole moments are, let us consider the dipole
moments of other species present in an actual experiment. Water has
a gas phase dipole moment of $\sim 1.8$ D~\cite{Clough1973-1}.
However, in the condensed phase, the dipole moment of water is
larger, $\sim 3$ D~\cite{Silve1999-1}.
%Could also ref. Gregory1997-1 where they give 2.4-2.6 D
This is easy to understand: the hydrogen
bonding that takes place between water molecules induces more
polarization in the electronic distribution, causing an increase in
the dipole moment. This is actually true for other species as well:
the interactions between the polar solvent and the solute (the PN) will
induce larger dipole moments in the solute. More specifically,
differences in how water interacts with the different bases could
significantly change their dipole moment (this is an open and
important question).

In addition, the ions present in solution contribute to the dipole
moment fluctuations. Consider, for instance, the counterion on the
DNA backbone. The values of the deoxyribonucleic acid dipoles given
in Table~\ref{dipoles} are for a passivated backbone. The
passivation is simply the addition of a hydrogen atom on one of the
partially charged oxygen atoms on the phosphate group. This
additional hydrogen has a bond length of about 1$\mbox{\AA}$.
Therefore, if we consider the effect of a bound counterion, we have
to add an additional dipole moment of about 1 $e \mbox{\AA} \;$,
because the ion-oxygen bond would be about 1 $\mbox{\AA}\;$ longer.
This additional dipole thus contributes 4.8 D. Further,
molecular dynamics simulations indicate that the counterions on the
DNA backbone fluctuate quite a lot~\cite{Lagerqvist2006-2}. Thus,
there could be noise larger in magnitude than the nucleotide or base
dipole moments themselves, and much larger than the differences
between them. However, how this noise interferes with the signal
from the bases is not clear, since at these length scales the
fluctuations may or may not give a small average noise.

\subsection{Optics}
\label{sec:seqoptic}

\begin{figure}
\begin{center}
\includegraphics*[width=5.5cm]{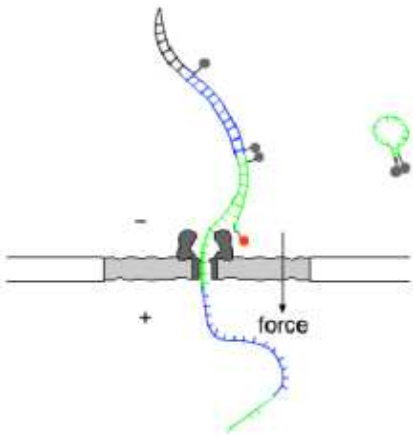}
\caption{(Color in online edition) A double strand of DNA with magnified sequence is pulled
through a nanopore. As the strand goes through, it unzips, releasing
the fluorescently tagged single strand, which is then read
optically. Courtesy of A. Meller.} \label{opt}
\end{center}
\end{figure}

Another proposed method for sequencing relies on optical methods
and ds-DNA unzipping via the interaction with a
nanopore.
The idea is to tear apart ``magnified'' DNA
and then read an optical signal~\cite{Lee2007-1}. This process is
shown in Fig.~\ref{opt}.

In more detail, one uses a biochemical process to create
``designer'' ds-DNA where each base is represented by a unique
$N$-base sequence (with $N\approx 20$). These $N$-base sequences are
just two blocks of $N/2$ base homogeneous strands. The $N/2$ base
strands are each labeled with a fluorescent tag on one end and
a quenching molecule at the other, such that within the double strand
each tag is always paired to a quencher. This prevents fluorescence
when the DNA is away from the pore. When the ds-DNA is
pulled into, for instance, an $\alpha$-hemolysin pore, the
double strand has to unzip into single strands~\cite{Mathe2004-1,Mathe2006-1,Sauer2003-1}.
This pulls the fluorescent
tag away from its quencher and allows the tag, and thus the base in the original
strand, to be detected via optical means.

Despite the necessity to work with ``amplified'' DNA,
%~\footnote{As
%previously discussed, amplification of DNA may introduce additional
%costs and errors in the sequencing procedure.}
this method has tremendous potential for sequencing, and can be
parallelized quite easily by creating several pores in the same
device~\cite{Lee2007-1}. Preliminary estimates give a raw sequencing
rate of $~\sim 1 - 10$ million bases/s for a parallelized nanopore
detector.

\subsection{Force}
\label{sec:seqforce}

\begin{figure}
\begin{center}
\includegraphics*[width=7.5cm]{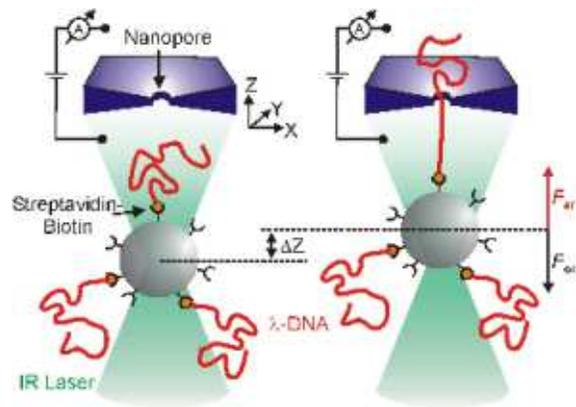}
\caption{(Color in online edition) Optical tweezer used in conjunction with a nanopore.
From~\onlinecite{Keyser2006-1}.}
%copyright Nature Publishing Group
\label{opttweez}
\end{center}
\end{figure}

One important technique that may be useful in DNA sequencing and
detection, and also single-molecule and nanopore studies, is the use
of optical tweezers~\cite{Neuman2004-1}. An already implemented
setup using optical tweezers in conjunction with a nanopore is shown
in Fig.~\ref{opttweez}~\cite{Keyser2006-1}.

In this experiments, $\lambda$-DNA is attached to polystyrene
beads via a streptavidin-biotin interaction
(this is non-covalent protein-ligand association, and in this case
is very strong~\cite{Diventra2004-1}). The dielectric bead can be trapped at the focal
point of a tightly focused laser beam~\cite{Neuman2004-1}. If the
bead is pulled away from the focal point, a restoring force is
exerted on the bead. For small displacements this force is linearly
proportional to the distance of the bead from the focus, e.g., it
resembles a spring obeying Hooke's law. Thus, by measuring the
position of the bead from the laser focus, which can be done by
measuring the reflected light off the bead, one can determine the
net external force on the bead via
\begin{equation}
F_{ot}=-k_{trap}\Delta Z
\end{equation}
where $k_{trap}$ is the stiffness of the optical trap and $\Delta Z$
is the distance of the bead from the focus of the
beam~\cite{Keyser2006-1}.

By bringing the bead close to a nanopore, one can then both control
the motion of the DNA in the nanopore and measure the forces exerted
on it by the pulling voltage. By measuring these forces, \onlinecite{Keyser2006-1} 
have determined that the effective charge for each
nucleotide pair on the double-stranded DNA is about 25$\%$ of its bare
value, i.e., 0.5 $e$ instead of 2 $e$. For 100 mV pulling voltage,
this effective charge gives a force on the DNA of 24 pN.

Clearly, the concept of coupling the nanopore setup with an optical
trap would open up many possibilities to study physics and chemistry
of single molecules. One can also imagine many possibilities for
using this setup, or more complex setups (for instance with two
optical traps, one on each side of the pore), for sequencing
technology and also to help in proof-of-principle prototypes. \onlinecite{Keyser2006-1} 
have succeeded in slowing the translocation of ds-DNA
down by five orders of magnitude compared to its free translocation
rate -- bringing it down from $\sim 8$ mm/s to 30 nm/s. For ionic blockade sequencing, the
ability of the optical trap to slow and control the DNA
translocation would allow for more definitive experimental
measurements of the different ionic blockade currents for the
various bases, and may also allow for the stretching of DNA so that
secondary structure effects may be reduced or removed.

For transverse current and capacitance
measurements, the controlled motion of DNA in the nanopore can also
help in a number of ways. For one, it can slow down the DNA so that
one can see the gaps between the bases and understand how to count
the bases as they go by the pore. If one were to use a more complex
setup, with two optical traps on each side of the pore, the
pulling voltage could be completely turned off. This would enable
the condition $E_{\parallel} \ll E_{\perp}$ necessary for successful
transverse transport sequencing to be satisfied (see
Sec.~\ref{sec:seqcurr}). Furthermore, it would reduce ion flow
through the pore to a minimum (only Brownian type motion would be
present) and may thus reduce noise from ionic fluctuations for both
the transverse current and capacitance.

In addition to the measurement of the effective charge of DNA in
solution, optical traps have provided the necessary control and
precision force measurements for the study of DNA interacting with
RNA polymerase at the single-base level~\cite{Abbondanzieri2005-1}
and even single-molecule sequencing~\cite{Greenleaf2006-1}.

%For the latter, four aspects of
%this procedure can be employed: the sub-nanometer position
%resolution of the optical tweezers, known sequences of DNA used to
%align different ``records'',
%four records each corresponding to each nucleotide (used by the RNA
%polymerase to synthesize the RNA complementary to the DNA) in low
%concentrations, and a histogram analysis. The rate at which DNA can be sequenced with this
%method is, however, limited by the RNA synthesis rate, which is
%several nt/s~\cite{Greenleaf2006-1}.

\section{Conclusions}
\label{sec:conclu}

In this Colloquium we have examined a variety of different proposals
for accurate, rapid DNA/RNA sequencing and detection. These methods
take {\it physical approaches} to detecting single bases in a
sequence. Some of the basic ideas have already been implemented
experimentally and shown to be useful for detection of strands of
DNA and also as probes for ``global'' properties, such as length or
homogeneous sequences. It is, however, still unclear as to whether
these results can be extended to true single-molecule sequencing of
DNA.

We showed explicitly with ionic blockade and transverse transport
 that the physical
differences of base properties come in the form of differing signal
{\em distributions}. This will hold true for all other approaches
because of the electronic and structural similarity of the bases. Thus,
a single, instantaneous measurement signal of a microscopically
fluctuating nucleotide can not be expected to fall within a unique
range. This suggests a
statistical approach to physical DNA sequencing. In some cases, the
statistics may be built into the measurement apparatus via time
averaging of the signals, as is the case with finite bandwidth electrical
measurements. It is still an open question
whether noise, in particular ionic noise, will drown out any
signal difference between the bases, especially with capacitance or transverse
transport measurements. Other noise has to be reduced in the inherently noisy
nanopore (or other) environment, and the measurement device has to
have sufficient resolution to see the differences in base signals.

Another open issue involves achievable read-lengths and sequencing rates.
Various lengths of DNA have been pulled through nanopores. For
instance,~\onlinecite{Storm2005-1} pulled through $\sim$100 kbp ds-DNA
through a 10 nm pore. However, the authors do not know of a
systematic study of possible read-lengths of ss-DNA being pulled
through nanometer scale pores. Since the forces on DNA in the pore are small (on the order of 10's of piconewtons, much
less than the forces necessary to break covalent bonds),
practical matters like entanglement of ss-DNA outside
the pore or nonlinear dependence of translocation velocities on DNA length are
likely to limit read-lengths.
Additionally, the translocation rate of each DNA through the nanopore
has to be smaller than the device bandwidth, allowing each base to stay long enough at its optimal
position in the detection apparatus (e.g., in the active detection
region, like between the two electrodes in a transverse transport
approach).

Progress in techniques and proof of principle is likely to proceed in part via the use of auxiliary
systems. Optical tweezers~\cite{Keyser2006-1}, a nanotube bound
to an AFM~\cite{King2005-1}, and other ideas~\cite{Krieger2006-1}
could give rise to techniques to slow polynucleotide (or molecular)
motion in the pore, and also offer the potential to get rid of the
ionic solution/electrophoretic pulling. This could be
tremendously beneficial in demonstrating proof of concept and/or
noise reduction. Further, we alluded to the distinction between chemical and physical
processes earlier. Chemical processes change the atomic makeup or
structure of a species. They potentially add slow extra steps to a
sequencing procedure. However, they can add enhanced
distinguishability as well. For instance, one could imagine using
the amplification of the optical method together with ionic blockade
or transverse transport. Since the nanopores can detect differences
in homogeneous sequences, it may be possible to detect amplified or
designer blocks of nucleotides.

In addition to the technological motivation, there is also a scientific motivation
for studying the detection step of a sequencing protocol. The process of directly detecting
physical differences between the bases is fundamentally different
from the chemical/optical processes currently in use. Thus, not only
does it represent a scientific challenge with the goal of
low-cost and rapid sequencing and detection, it is likely
to increase our understanding of biological and other molecules
at the atomic scale, and may lead to off-shoot
technologies.
%might could put that nanoscale sensors and dectors of DNA/RNA and other
%molecules might enable to the study of the role of fluctuations and localization in
%biology and chemistry. One example is Block's work on polymerase. I believe via their
%optical trap experiments they were able to show that the polymerase didn't operate
%with a rachet mechanism (or maybe I have it backwards) ... which is a good example
%of fundamental science that can be explored with these techniques.

In this Colloquium, we have raised several important open
questions. Now that fabrication techniques of solid state pores are
maturing, experiments can be tailored to answer these questions more
directly by systematically changing pore dimensions, materials,
surface properties, and environmental (e.g., ionic) conditions.
With adequately controlled experiments, nanopores could be used to
probe inhomogeneities at the nanoscale, such as fine structures in ``dielectrics''
(e.g., hydration layers) and other interesting physical
phenomena.

Because of the widespread technological and scientific interest, it
is impossible to predict where this field will go from here.
Undoubtedly, ingenious techniques and methodologies will be invented
which extend our abilities to detect and sequence DNA well beyond
what has been discussed. Some examples have been mentioned:
integration of biological pores with nanoscale probes, use of
molecular base discriminators with biological or synthetic pores,
etc.
The most promising fundamental research will sort
through the many competing effects that exist in these complex
systems, and increase our understanding of physical processes at the
interface between solids, liquids, and biomolecules down to the
nanometer scale regime.
 We are confident that a physical approach to DNA detection will yield a wealth
of information on physical and chemical processes at the nanoscale,
and, hopefully, a rapid and low-cost sequencing technology.

\section*{Acknowledgments}

We would like to thank M. Ramsey for useful discussions, J.
Lagerqvist and N. Bushong for a critical reading of the manuscript, and A. Meller
for providing us with information about his suggested sequencing
method before publication. MZ acknowledges support from
a Gordon and Betty Moore Fellowship and MD acknowledges support from the National Human Genome
Research Institute of NIH, Grant \# 2 R01 HG002647-03.

\end{document}